\documentclass[final,11pt,times,authoryear]{elsarticle}
\usepackage[table]{xcolor}
\usepackage{lineno}
\usepackage[colorlinks = true, linkcolor = blue, urlcolor = blue, citecolor = blue, anchorcolor = blue]{hyperref}
\usepackage{diagbox}
\usepackage{multirow}
\usepackage{fancyhdr}
\usepackage{amsmath}
\usepackage{amssymb}
\usepackage{indentfirst}
\usepackage{booktabs}
\usepackage{graphicx}
\usepackage{epstopdf}
\usepackage{bm}
\usepackage{float}
\usepackage{subfigure}
\usepackage[section]{placeins}
\usepackage{geometry}
\usepackage{flushend,cuted}
\usepackage{caption}
\usepackage{booktabs}
\usepackage{array} % For custom column width definitions (P)
\newcolumntype{P}[1]{>{\centering\arraybackslash}p{#1}} % Define P column type for centered text in fixed-width columns
\usepackage{xcolor,soul}
\definecolor{darkgreen}{rgb}{0.0,0.5,0.0}
\definecolor{lavender}{rgb}{0.5,0.25,0.5}
\definecolor{darkred}{rgb}{0.85,0.0,0.15}
\makeatletter

\newcommand{\Rmnum}[1]{\expandafter\@slowromancap\romannumeral #1@}
\makeatother

\journal{International Journal of Plasticity}
\makeatletter
\def\ps@pprintTitle{%
 \let\@oddhead\@empty
 \let\@evenhead\@empty
 \def\@oddfoot{\footnotesize\hfill\thepage\hfill}
 \let\@evenfoot\@oddfoot}
\makeatother
\begin{document}

\begin{frontmatter}

\title{\textbf{Atomistic and data‑driven insights into the local slip resistances in random refractory multi-principal element alloys}}

\author[1,2]{Wu-Rong Jian\footnote{Corresponding author. Email address: {\tt wurong@scut.edu.cn} (Wu-Rong Jian)}}

% \corref{corr-author}} \ead{wurong@scut.edu.cn}

\author[3]{Arjun S.\ Kulathuvayal}

\author[4]{Hanfeng Zhai}

\author[3]{Anshu Raj}

\author[1,2]{Xiaohu Yao}

\author[3]{Yanqing Su}

\author[3]{Shuozhi Xu}

\author[5,6]{Irene J.\ Beyerlein}

\address[1]{State Key Laboratory of Subtropical Building Science, South China University of Technology, Guangzhou, Guangdong 510640, P. R. China}
\address[2]{Department of Engineering Mechanics, South China University of Technology, Guangzhou, Guangdong 510640, P. R. China}
\address[3]{School of Aerospace and Mechanical Engineering, University of Oklahoma, Norman, OK 73019-1052, USA}
\address[4]{Department of Mechanical Engineering, Stanford University, Stanford, CA 94305, USA}
\address[5]{Department of Mechanical Engineering, University of California, Santa Barbara, CA 93106-5070, USA}
\address[6]{Materials Department, University of California, Santa Barbara, CA 93106-5050, USA}

\date{\today}

\begin{abstract}\small
Refractory multi-principal element alloys (RMPEAs) have garnered considerable interest for their exceptional high-temperature strength and promising applications in demanding environments. However, the differing and complex compositions among many RMPEAs pose significant challenges to understanding the deformation mechanisms that govern their plastic deformation. In this work, we address these challenges by conducting atomistic simulations to determine the local slip resistances (LSRs) for edge and screw dislocations on the \{110\}, \{112\}, and \{123\} slip planes in 12 body-centered cubic (BCC), equal-molar RMPEAs. To elucidate the relationship between LSR and the underlying physical properties of the alloys, we employ machine learning methods, enabling a systematic analysis of how compositional variations affect the dislocation behavior. Building on these insights, we develop an analytical model based on thermally activated theory to predict the macroscopic yield stress of RMPEAs. Our results demonstrate that increasing the fraction of hexagonal close-packed (HCP) elements above 50\% in the alloy composition notably reduces both the unstable stacking fault energy (USFE) and ideal shear strength (ISS), further lowering the LSRs of screw dislocations on all three slip planes. Furthermore, higher elastic anisotropy (quantified by the Zener ratio) diminishes USFE, ISS, and LSR, while severe lattice distortion (LD) reduces the screw-to-edge LSR ratio but enhances the \(\{110\}\)-to-\(\{112\}\) and \(\{110\}\)-to-\(\{123\}\) LSR ratios. By integrating an autoencoder with a random forest model, we analyze the correlation between material properties and LSR, revealing that elastic constants and LD coefficients play the most critical roles in determining LSR. Our thermally activated, dislocation-based model, which integrates the contributions of both edge and screw dislocations on all three slip planes, provides reliable predictions of tensile yield stress for BCC RMPEAs. This framework offers a robust foundation for alloy design and optimization in high-performance applications.
\end{abstract}


\end{frontmatter}

\clearpage
%%%%
%%%%
\section{Introduction}
\label{sec:introduction}

Compared to conventional alloys composed of a single dominant element, body-centered cubic (BCC) refractory multi-principal element alloys (RMPEAs), containing three or more principal refractory elements, have garnered significant attention from materials scientists due to their superior mechanical properties \citep{senkov2018dejmr}, such as high yield strength at elevated temperatures \citep{wei2024ijp,zhang2025mc}, large ultimate strain \citep{wang2024ijp} and significant strain hardening \citep{wang2025ijp} at high strain rate, and excellent resistance to irradiation \citep{mo2024ijp} and creep \citep{feng2024ijp}. Further enhancement of the mechanical performance of RMPEAs necessitates a deeper understanding of their plastic deformation mechanisms. Since dislocations serve as the primary carriers of plasticity in BCC alloys \citep{anderson2017theory,liu2024ijp}, a comprehensive understanding of dislocation slip behavior in RMPEAs is essential for developing strategies to design alloys with enhanced yield strength.

For dislocation slip, it is crucial to determine the critical resolved shear stress (CRSS) at which the energy barrier for dislocation movement becomes zero, allowing a dislocation to glide on a specific plane. In pure metals, the CRSS at 0~K is known as Peierls stress \citep{peierls1940pps}. Compared to pure metals, RMPEAs contain multiple elements of different atomic sizes, exhibiting inevitable lattice distortion (LD) \citep{lee2020am,wang2024nc}. In such chemically and structurally heterogeneous materials, the traditional concept of a spatially constant Peierls stress is no longer applicable. Instead, the local resistance to dislocation glide varies from one atomic site to another due to fluctuations in chemical composition and lattice strain. Molecular dynamics simulations have elucidated the critical role of LD in modulating the deformation mechanisms of MPEAs. Specifically, LD has been shown to facilitate the nucleation of dislocations, as demonstrated by Jian et al. \citep{jian2020acta}, who found it lowers the strain required for Shockley partial dislocation nucleation in CoCrNi, and by Hua et al. \citep{hua2023ijp}, who observed its potent effect on promoting dislocation nucleation at grain boundaries. Furthermore, LD can also influence phase stability, with Wu et al. \citep{wu2023ijp} reporting that it aids a FCC-BCC transformation while simultaneously hindering the nucleation of stacking faults in a AlCuCoFeNi MPEA at low temperatures. Furthermore, the heterogeneous energy landscape created by LD forces dislocations to bend locally, navigating paths of least resistance and resulting in characteristically wavy dislocation lines \citep{li2019nc,smith2020ijp,zheng2023npjcm}. Unlike pure metals, where the energy barrier for dislocation motion remains uniform along a straight dislocation line of identical character, the presence of LD in RMPEAs leads to variations in the energy barrier along the wavy dislocation line. Consequently, Peierls stress, which is constant along the dislocation line in pure metals, is assumed to vary along the wavy dislocation line in RMPEAs. This variation arises because the local atomic environment, governed by the specific arrangement of constituent elements and the associated LD, modifies the underlying periodic potential that a dislocation must overcome.

To better characterize the 0~K CRSS in RMPEAs and differentiate it from Peierls stress in pure metals, the concept of local slip resistance (LSR) has been introduced \citep{wang2020science,xu2021acta}. Like the Peierls stress, the LSR is the 0~K critical shear stress to move a straight dislocation line segment (approximately 1-2~nm) in the crystal on a given slip plane. However, whereas the Peierls stress is a single material constant, the LSR is defined locally and statistically: for a given alloy composition, the LSR is calculated at many different locations within the disordered structure, each representing a distinct local atomic configuration. The resulting distribution of LSR values captures the spatial heterogeneity of the glide resistance. To represent the resistance to glide in a RMPEA with a specific composition, one can, for instance, take the average of the LSR distribution. Importantly, the LSR does not exist independently of the Peierls stress; rather, it generalizes the Peierls concept to heterogeneous materials. Each local LSR value can be understood as a locally modified Peierls stress, where the modification stems from the local chemical composition and the accompanying LD. Factors such as alloying segregation alter the local atomic environment, thereby directly affecting the local energy barrier for dislocation glide. These effects are not simply additive to a baseline “pure-metal” Peierls stress; instead, the LSR embodies the combined outcome of the intrinsic lattice potential and its perturbation by chemical and structural disorder. Thus, the LSR provides a comprehensive descriptor that inherently accounts for the spatially varying slip resistance in RMPEAs.

In BCC pure metals, yielding is primarily governed by screw dislocations that exhibit Peierls stresses several times higher than those of edge dislocations on the same glide plane \citep{wang2021cms}. Recent studies have calculated the LSRs of various BCC RMPEAs, including a ternary alloy (MoNbTi) \citep{wang2020science,xu2021acta}, five quaternary alloys (CrMoNbTa, CrNbTaW, MoNbTaV, MoNbTaW, NbTaTiV) \citep{romero2022ijp,nitol2024cms}, and two quinary alloys (NbTaTiVZr, MoNbTaVW) \citep{nitol2024cms,wang2024npjcm}. These investigations examined the LSRs of both edge and screw dislocations on the most common glide planes, such as \{110\}, \{112\}, and \{123\}. For a given glide plane in a given RMPEA, the screw-to-edge LSR ratio typically ranges from 1 to 2 \citep{romero2022ijp}, significantly lower than the corresponding Peierls stress ratios ranging from one to two orders of magnitude in BCC pure metals, whose elements constitute these RMPEAs \citep{kang2012pnas,wang2021cms}. This finding suggests that edge dislocations play a role comparable to screw dislocations in the yielding of BCC RMPEAs, a conclusion supported by experimental observations in the MoNbTi RMPEA \citep{wang2020science}. Among these RMPEAs, six BCC elements are present, with Cr having a notably smaller atomic radius (128 pm) than the other five BCC elements: Mo (140 pm), Nb (147 pm), Ta (147 pm), V (135 pm), and W (141 pm) \citep{zhang2015calphad}. This significant atomic-size mismatch between Cr and the other BCC elements leads to an pronounced increase in the LD. In addition, the inclusion of Cr atoms in RMPEAs (such as CrMoNbTa and CrNbTaW) directly results in a relatively higher average LSR and a more dispersed distribution of LSR values \citep{romero2022ijp}.

Furthermore, dislocation slip in these alloys involves the breaking and reforming of atomic bonds across the glide plane, a process analogous to shearing or displacing one half of the crystal relative to the other along a specific plane \citep{anderson2017theory}. The energy cost per unit area during this process is represented by the generalized stacking fault energy (GSFE), often visualized as a curve showing the energy variation with the relative displacement between the two halves of the crystal \citep{vitek2011pms}. The ideal shear strength (ISS) of the alloy can be derived from the maximum slope of the GSFE curve, corresponding to the maximum shear stress during displacement along the glide plane.

Despite these advancements, the findings are limited to a small number of BCC RMPEAs, raising questions about their general applicability. Additionally, metallic elements with hexagonal close-packed (HCP) crystal structures, such as Ti and Hf, are increasingly incorporated into BCC RMPEAs to improve ductility \citep{zhang2023ijp,tsuru2024nc,wang2024ijp,zhang2025ijp}. However, the effect of these HCP elements on the LSR of BCC RMPEAs remains poorly understood. Beyond LD and ISS, other fundamental physical properties, such as lattice parameters, elastic constants, cohesive energy, and unstable stacking fault energy (USFE), may also influence the LSR values of edge and screw dislocations on various glide planes. The extent to which these intrinsic properties impact LSR values remains unclear. Furthermore, the relationship between atomic-scale LSR and macroscopic yield strength warrants systematic investigation to better elucidate the plastic deformation mechanisms of BCC RMPEAs across a wide range of temperatures.

In this work, molecular static (MS) simulations are conducted to calculate the LSR values on the \{110\}, \{112\}, and \{123\} slip planes in a total of 12 BCC equal-molar RMPEAs, including nine ternaries, one quaternary, and two quinaries. Machine learning (ML) methods are then employed to analyze the correlation between the LSR values and the fundamental physical properties of various RMPEAs. Following this, an analytical model, grounded in thermally activated theory, is developed to predict the macroscopic yield stress of BCC RMPEAs. It is found that increasing the fraction of HCP elements in the RMPEA composition above 50\% significantly lowers both USFE and ISS, which also causes low LSR values for screw dislocations on all three slip planes in the Hf- or Ti-containing alloys. Additionally, higher elastic anisotropy, as indicated by a larger Zener ratio, also diminishes USFE, ISS, and LSR. Severe LD lessens the screw-to-edge LSR ratios but elevates the \(\{110\}\)-to-\(\{112\}\) and \(\{110\}\)-to-\(\{123\}\) LSR ratios in these alloys. By combining an autoencoder with a random forest model, we uncover the relationship between material properties and LSR, identifying elastic constants and LD coefficients as the most influential factors governing LSR. The built thermally activated, dislocation-based model, incorporating both edge and screw dislocations on \{110\}, \{112\}, and \{123\} planes, provides robust yield stress predictions for BCC RMPEAs.

\section{Methodology}
\label{sec:methodology}

%%%Fig1
\begin{figure}[htbp]
\centering
\includegraphics[scale=0.5,clip]{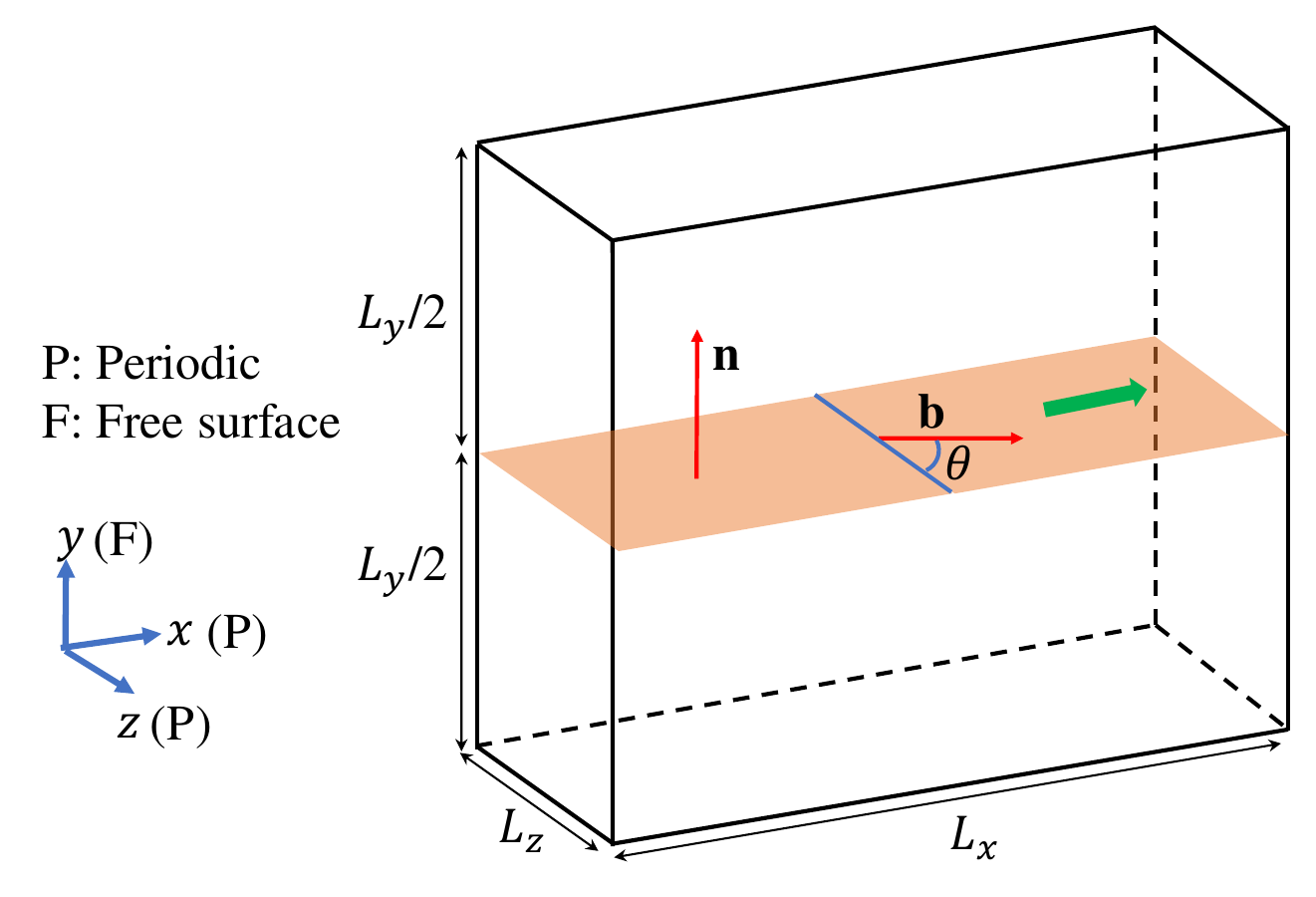}
\caption{The schematic for local slip resistance (LSR) calculations. The dislocation line, its glide plane and Burgers vector on this plane are denoted by blue line, light-orange parallelogram and red arrow, respectively. The dislocation line and the normal of its glide plane are along $z$ and $y$ axes, respectively. For edge dislocation, the character angle $\theta$ between dislocation line and Burgers vector is $90^\circ$, while for screw dislocation, $0^\circ$.}
\label{fig:1}
\end{figure}

All simulations in this work are conducted via Large-scale Atomic/Molecular Massively Parallel Simulator (LAMMPS) \citep{thompson2022cpc}, where 12 RMPEAs are considered, i.e., NbTaTi, MoNbTi, HfNbTa, NbTiZr, HfNbTi, HfTaTi, TaTiZr, MoTaTi, MoNbTa, HfNbTaTi, HfMoNbTaTi and HfNbTaTiZr. The interactions among the six types of constituent atoms (Hf, Mo, Nb, Ta, Ti, and Zr) in these alloys are described using an embedded-atom method (EAM) alloy potential, created recently by \citep{mubassira2025cms} using the strategy proposed by \citep{zhou2004prb}. For this EAM potential, cross-terms between different atomic species are built by averaging the single-elemental potentials, based on the formulations of \citep{johnson1989prb}. This approach for developing alloy potentials has been applied widely in the previous studies \citep{maresca2020acta1,xu2022cms}.  

A single-crystal BCC configuration consisting of identical atoms, with the [100], [010], and [001] orientations aligned along the $x$, $y$, and $z$ axes respectively, is constructed, containing 10 lattices along each dimension. To achieve the desired equal-molar composition, some atoms are then randomly replaced by other atomic species. Subsequently, some fundamental properties of the RMPEAs, including lattice constant ($a_{0}$), elastic constants ($C_{11}$, $C_{12}$, $C_{44}$), Zener ratio ($A_{\rm c} = 2C_{44}/(C_{11}-C_{12})$), and cohesive energy ($E_{\rm coh} = E_{\rm total}/N_{\rm atom}$), are determined using MS simulations. The LDs of RMPEAs were quantified using the full width at half maximum (FWHM) of the radial distribution functions $g(r)$ that is defined as the first nearest neighbor shell over the interatomic distance $r$ \citep{jian2020acta}. Here, we define the LD coefficient as $\delta = {\rm FWHM}/a_{0}$. A similar but larger single-crystal BCC configuration, with 50 lattices along each dimension, is also built as a reference to obtain some of the same properties. 

%%%Table1
\begin{table}[htb!]
\centering
\caption{The crystallographic orientations of atomic configurations used in the LSR calculations for \{110\}, \{112\}, and \{123\} slip planes.}
\label{tbl:cry_ori}
\begin{tabular}{ccccc}
\hline
& & \{110\} & \{112\} & \{123\} \\ \hline
& $x$ & [111] & [111] & [111] \\
Edge & $y$ & $[1\bar{1}0]$ & $[\bar{1}\bar{1}2]$ & $[\bar{1}\bar{2}3]$ \\
& $z$ & $[11\bar{2}]$ & $[1\bar{1}0]$ & $[5\bar{4}\bar{1}]$ \\ \hline
& $x$ & $[\bar{1}\bar{1}2]$ & $[\bar{1}10]$ & $[\bar{5}41]$ \\
Screw & $y$ & $[1\bar{1}0]$ & $[\bar{1}\bar{1}2]$ & $[\bar{1}\bar{2}3]$ \\
& $z$ & [111] & [111] & [111] \\ \hline
\end{tabular}
\end{table}

The LSR calculations are based on another single-crystal BCC configuration with the dimensions of 50 nm ($L_{x}$) $\times$ 50 nm ($L_{y}$) $\times$ 1$\sim$2 nm ($L_{z}$). An edge or screw dislocation is inserted at the center of the $xy$ cross-section. As shown in \autoref{fig:1}, the $x$, $y$ and $z$ axes are along the glide direction of dislocation, the normal direction of the slip plane, and the line direction of dislocation, respectively. Only edge and screw dislocations gliding on the \{110\}, \{112\}, and \{123\} planes are considered for the LSR calculations. For atomic configurations involving various slip planes, the $x$, $y$, and $z$ axes are oriented along different crystallographic orientations, as detailed in \autoref{tbl:cry_ori}. During the LSR calculations, periodic boundary conditions (PBCs) are applied along the $x$ and $z$ directions, while traction-free boundaries are set along the $y$ direction. This configuration, known as the periodic array of dislocations (PAD), has been widely used in previous atomistic simulations involving Peierls stress \citep{jian2021cms,kang2012pnas}, dislocation mobility \citep{cho2017ijp,lunev2018ijp,jian2020msmse}, and LSR \citep{wang2020science,xu2021acta,romero2022ijp}. Two atomic layers near the top and bottom surfaces along the $y$ direction are regarded as boundary layers. To initiate dislocation movement, the simulation cell is deformed by applying an incremental strain tensor $\mathrm{\Delta} \bm{\varepsilon}$ with only one non-zero component. For edge dislocation, $\mathrm{\Delta} \bm{\varepsilon}$ is set to $|\mathrm{\Delta} \varepsilon_{yx}| = 10^{-5}$, while for screw dislocation it is set to $|\mathrm{\Delta} \varepsilon_{yz}| = 10^{-6}$. The remaining components of $\mathrm{\Delta} \bm{\varepsilon}$ are kept zero. These settings have been verified in the previous works to be sufficiently small to avoid unreliable results \citep{xu2021acta,romero2022ijp}. Following each strain increment, the configuration is relaxed using the conjugate gradient scheme and the fast inertial relaxation engine \citep{xu2021acta}. During energy minimization, atoms in the boundary layers are restricted from moving along the $x$ direction for edge dislocations and the $z$ direction for screw dislocations, while all other atoms are allowed to relax along any direction. With increasing strain, the dislocation glides. The minimum shear stress required to move the dislocation at least 1 nm within one strain increment is defined as the LSR ($\tau_\mathrm{lsr}$). The choice of a relatively short dislocation length (1-2 nm) in this work is much shorter than the typical critical length for kink-pair nucleation in pure BCC metals (often on the order of $\sim$10 nm \citep{hirth1982theory}). However, the fundamental objective here differs: in RMPEAs, the LSR is explicitly defined as the critical shear stress required to move a straight dislocation segment locally within a specific local atomic environment, deliberately isolating the intrinsic lattice friction from the dynamics of kink-pair formation and propagation. This approach is standard in atomistic investigations of LSR, as evidenced by prior studies on similar systems \citep{wang2020science,xu2021acta,romero2022ijp,wang2024npjcm,nitol2024cms}. By employing this shorter segment, we directly probe the atomistic-scale resistance arising from chemical heterogeneity and lattice distortion, which is the core mechanism of interest. This setup allows us to generate the statistical distribution of local energy barriers without the complicating factors of long-range line tension effects or kink-mediated motion, providing a clear baseline for understanding the role of compositional complexity on dislocation glide. For statistical accuracy, all calculations of the above properties and LSRs are performed for 10 samples with different random atomic distributions for each RMPEA composition, and the averaged values are reported. For comparison, the Peierls stresses of BCC Mo, Nb, and Ta are also calculated.

Additionally, we calculate the GSFE curves for the $\{110\}$, $\{112\}$, and $\{123\}$ planes, using cubic atomic configurations with an edge length of 27 nm. Their crystallographic orientations are consistent with those used for LSR calculations of edge dislocations, as shown in \autoref{tbl:cry_ori}. During the calculations of GSFE curves, PBCs are only applied along the $x$ and $z$ directions. The upper half of each sample is displaced by $d$ relative to the lower half along the [111] orientation ($x$-axis), and the GSFE curve is obtained as a function of $d$ from 0 to {$\sqrt{3}a_{0}/2$. More details on the GSFE calculations can be found in our previous studies \citep{romero2022ijp}. Each GSFE curve exhibits a single peak, whose value represents the USFE ($\gamma_{\mathrm{usf}} = \max[\gamma_{\mathrm{gsf}}(d)]$). From the GSFE curve, we determine the ISS ($\tau_{\mathrm{iss}} = \max[\partial \gamma_{\text{gsf}}(d)/\partial d]$), which is derived from the maximum slope of the GSFE curve, corresponding to the maximum shear stress during movement of the upper half of each sample \citep{xu2022aplm}. To ensure a good statistical representation, the GSFE curve, $\gamma_{\mathrm{usf}}$, and $\tau_{\mathrm{iss}}$ are averaged over 20 samples for each RMPEA composition.

\section{Results and discussion}
\label{sec:results}

\subsection{Fundamental properties of RMPEAs}
\label{sec:fundamental_properties}

To investigate the relationship between the fundamental properties of BCC RMPEAs and their LSR values, a series of fundamental properties of BCC RMPEAs are calculated. These properties include the LD coefficient ($\delta$), lattice parameter ($a_0$), elastic constants ($C_{11}$, $C_{12}$, $C_{44}$), Zener ratio ($A_{\rm c}$), cohesive energy ($E_{\rm coh}$), USFE for the \{110\}, \{112\}, and \{123\} planes ($\gamma^{110}_{\rm usf}$, $\gamma^{112}_{\rm usf}$, $\gamma^{123}_{\rm usf}$), and ISS for the same planes ($\tau^{110}_{\rm iss}$, $\tau^{112}_{\rm iss}$, $\tau^{123}_{\rm iss}$). The average values of these properties are summarized in \autoref{tbl:mech_sum}.

%%%Table2
\begin{table*}[]
\rotatebox{90}{
\begin{minipage}{\textheight} % Keep the table aligned and inside page margins
\caption{Mechanical properties of 12 RMPEAs from MS calculations. Lattice distortion coefficient: $\delta$; Lattice parameter (\AA): $a_0$; Elastic constants (GPa): $C_{11}$, $C_{12}$, $C_{44}$; Zener ratio: $A_{\rm c}$; Cohesive energy (eV): $E_{\rm coh}$; USFE ($(\mathrm{mJ}/\mathrm{m}^2)$): $\gamma^{110}_{\rm usf}$, $\gamma^{112}_{\rm usf}$, $\gamma^{123}_{\rm usf}$; ISS (GPa): $\tau^{110}_{\rm iss}$, $\tau^{112}_{\rm iss}$, $\tau^{123}_{\rm iss}$.}
\label{tbl:mech_sum}
\centering
\resizebox{\textwidth}{!}{
\begin{tabular}{c|ccccccccccccc}
\hline
\diagbox{Composition}{Values}  & $\delta$ & $a_0$ & $C_{11}$ & $C_{12}$ & $C_{44}$ & $A_{\mathrm{c}}$ & $E_{\mathrm{coh}}$ & $\gamma^{110}_{\mathrm{usf}}$ & $\gamma^{112}_{\mathrm{usf}}$ & $\gamma^{123}_{\mathrm{usf}}$ & $\tau^{110}_{\mathrm{iss}}$ & $\tau^{112}_{\mathrm{iss}}$ & $\tau^{123}_{\mathrm{iss}}$ \\
\hline
NbTaTi      & 0.0126 & 3.295 & 196.24 & 124.39 & 66.37 & 1.85 & $-6.76$ & 559.34 & 645.25 & 635.36 & 6.99 & 8.13 & 7.99 \\
MoNbTi      & 0.0282 & 3.233 & 241.38 & 133.00 & 76.32 & 1.41 & $-6.43$ & 742.35 & 857.21 & 845.89 & 7.44 & 8.38 & 8.32 \\
HfNbTa      & 0.0371 & 3.383 & 201.27 & 132.08 & 68.54 & 1.98 & $-7.39$ & 647.42 & 758.90 & 745.44 & 7.51 & 8.86 & 8.72 \\
NbTiZr      & 0.0429 & 3.391 & 143.32 & 110.55 & 56.85 & 3.47 & $-6.31$ & 459.61 & 537.78 & 529.89 & 5.17 & 6.03 & 5.95 \\
HfNbTi      & 0.0400 & 3.386 & 146.02 & 115.97 & 65.41 & 4.35 & $-6.29$ & 474.26 & 566.09 & 556.67 & 5.35 & 6.35 & 6.25 \\
HfTaTi      & 0.0356 & 3.381 & 154.69 & 119.64 & 82.03 & 4.68 & $-6.45$ & 511.04 & 610.78 & 599.62 & 5.81 & 7.03 & 6.93 \\
TaTiZr      & 0.0387 & 3.385 & 153.52 & 116.17 & 71.84 & 3.85 & $-6.46$ & 494.45 & 579.18 & 569.42 & 5.57 & 6.60 & 6.53 \\
MoTaTi      & 0.0254 & 3.240 & 234.35 & 140.46 & 92.70 & 1.97 & $-6.59$ & 786.30 & 910.26 & 897.48 & 8.45 & 9.87 & 9.76 \\
MoNbTa      & 0.0260 & 3.243 & 308.55 & 158.28 & 76.84 & 1.02 & $-7.57$ & 877.23 & 1016.40 & 1000.05 & 9.31 & 10.63 & 10.48 \\
HfNbTaTi    & 0.0354 & 3.361 & 175.31 & 123.25 & 70.57 & 2.71 & $-6.73$ & 553.96 & 653.34 & 641.87 & 6.46 & 7.62 & 7.53 \\
HfMoNbTaTi  & 0.0374 & 3.310 & 205.97 & 141.21 & 82.20 & 2.54 & $-6.82$ & 666.67 & 782.60 & 769.12 & 7.38 & 8.67 & 8.55 \\
HfNbTaTiZr  & 0.0420 & 3.405 & 157.62 & 118.17 & 67.61 & 3.43 & $-6.69$ & 522.79 & 615.94 & 605.31 & 5.82 & 6.88 & 6.80 \\
\hline
\end{tabular}
}
\end{minipage}
}
\end{table*}

\subsubsection{GSFE curves}
%%%Fig2
\begin{figure}[htp]
\centering
\includegraphics[scale=0.45,clip]{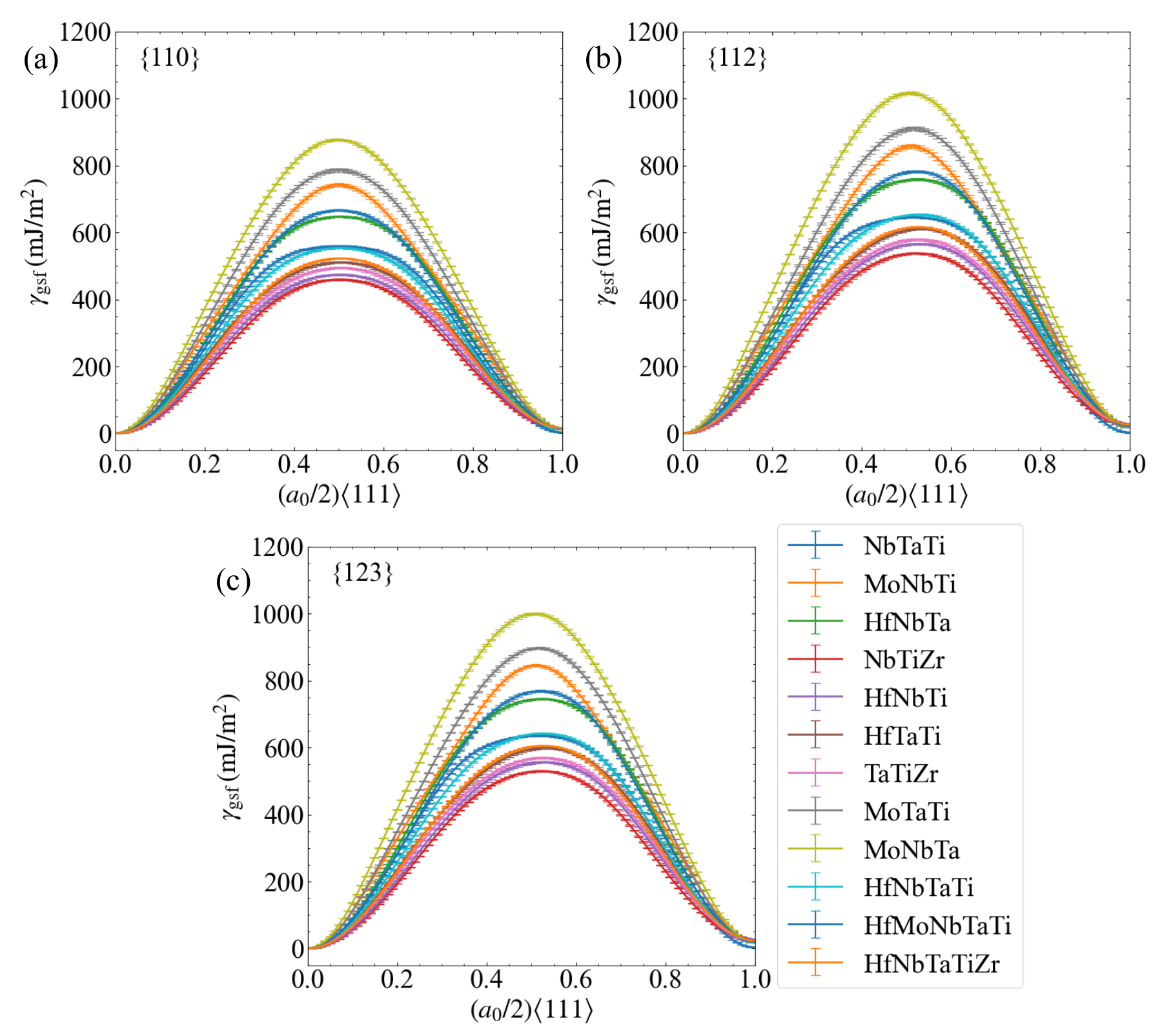}
\caption{The average GSFE curves of 12 RMPEAs on (a) $\{110\}$, (b) $\{112\}$, and (c) $\{123\}$ slip planes based on MS simulations. Each curve contains 101 data points; error bars represent the standard deviation derived from 20 independent samples per point.}
\label{fig:2}
\end{figure}

\autoref{fig:2} shows the average GSFE curves for 12 RMPEAs on the $\{110\}$, $\{112\}$, and $\{123\}$ slip planes. Each curve is calculated by performing relative displacements along the [111] orientation ($x$-axis) on the corresponding slip plane over a single periodic distance, which corresponds to the Burgers vector magnitude $b=\sqrt{3}a_{0}/2$. Within this periodic distance, $b$, all GSFE curves exhibit a single peak, with no local minima or metastable states observed. Unlike the perfectly symmetric GSFE curves of pure metals with respect to the peak, the GSFE curves of RMPEAs are asymmetric due to LD and chemical inhomogeneity \citep{romero2022ijp}. 

\subsubsection{USFE and ISS}
%%%Fig3
\begin{figure}[htp]
\centering
\includegraphics[scale=0.6,clip]{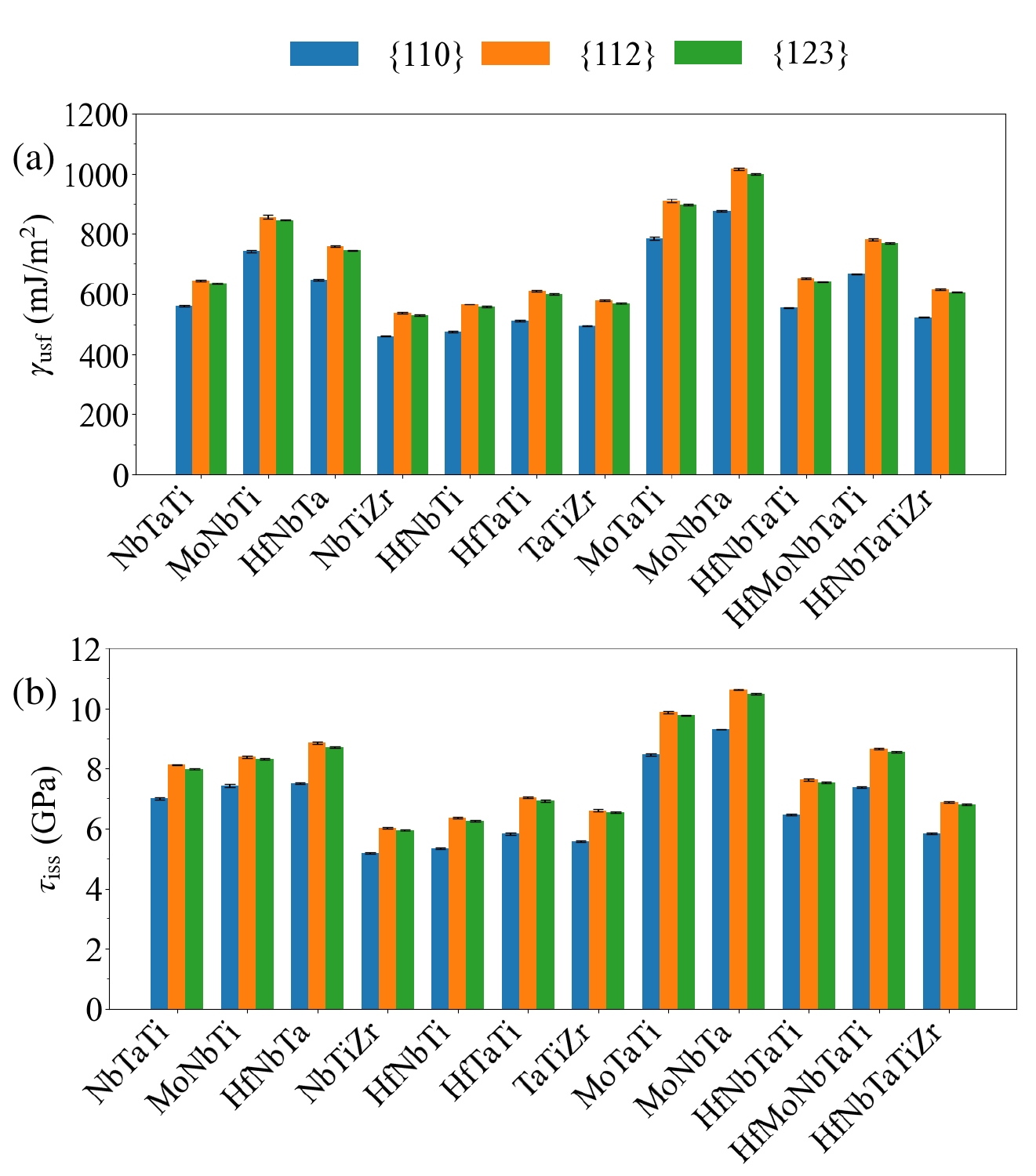}
\caption{The mean values of (a) USFE ($\gamma_{\mathrm{usf}}$) and (b) ISS ($\tau_{\mathrm{iss}}$) of 12 RMPEAs for $\{110\}$, $\{112\}$, and $\{123\}$ slip planes. Error bars represent the standard deviation derived from 20 independent samples per RMPEA.}
\label{fig:3}
\end{figure}

\autoref{fig:3}(a) and (b) presents the average  $\gamma_{\mathrm{usf}}$ and $\tau_{\mathrm{iss}}$ for all 12 RMPEAs on the $\{110\}$, $\{112\}$, and $\{123\}$ slip planes. Although both are theoretical quantities at 0~K and lacking dislocation core effects, the USFE and ISS can still serve as measures of resistance for dislocations to move under an applied stress. When both the USFE and ISS rise, dislocation glide becomes more difficult: a higher USFE elevates the energy cost for nucleation and motion, whereas a higher ISS implies the lattice endures larger shear stresses before yielding. Consequently, increases in both factors would translate to increases in the CRSS and hence are signatures of increased resistance to plastic deformation.

Among the 12 RMPEAs, the four highest USFE values on each slip plane consistently appear in the Mo-containing compositions, i.e., MoNbTa, MoTaTi, MoNbTi, and HfMoNbTaTi, listed in decreasing order (\autoref{fig:3}(a) and \autoref{tbl:mech_sum}). This indicates that the Mo element plays a significant role in enhancing the USFE value of RMPEAs, thereby possibly strengthening their resistance to dislocation glide. Mechanistically, this enhancement arises because Mo is a refractory metal with a strong directional bonding and a high intrinsic resistance to shear deformation. These characteristics stem from its high bond strength, which result in a large energy penalty for displacing adjacent atomic planes, i.e., a high intrinsic USFE. In pure Mo, the USFE values are reported as 1458.05 $\mathrm{mJ}/\mathrm{m}^2$ for \{110\} plane, 1689.03 $\mathrm{mJ}/\mathrm{m}^2$ for \{112\} plane, and 1657.93 $\mathrm{mJ}/\mathrm{m}^2$ for \{123\} plane \citep{xu2020intermetallics}. As shown in \autoref{tbl:USFE_ISS}, its USFE values are much higher than the other elements (Hf, Ti, Zr, Nb, Ta) considered in this work. When Mo is incorporated into an RMPEA, its local atomic environments retain this high resistance to shear. Consequently, regions rich in Mo atoms raise the average energy cost for the shear displacement across the slip plane. This effect is particularly pronounced in the Mo‑containing compositions studied (MoNbTa, MoTaTi, MoNbTi, and HfMoNbTaTi), where the high Mo content consistently leads to the highest USFE values among the 12 alloys. Thus, the enhancement of USFE in Mo‑containing RMPEAs is reflected by the large USFE of Mo.

\begin{table}[htb] 
\centering
\small
\caption{
USFE ($\mathrm{mJ}/\mathrm{m}^2$) and ISS (GPa) of BCC Hf, BCC Ti,  BCC Zr, BCC Mo, BCC Nb, and BCC Ta on the three types of slip planes calculated by MS simulations. The data for BCC Ti, BCC Zr, BCC Mo, BCC Nb, and BCC Ta are taken from our prior work \citep{xu2020intermetallics,wang2021cms}.
} 
\label{tbl:USFE_ISS}
\begin{tabular}{ P{3.5cm} P{1.3cm} P{1.3cm} P{1.3cm} P{1.3cm} P{1.3cm} P{1.3cm} }
\toprule
 Elements   & $\gamma^{110}_{\rm usf}$ & $\gamma^{112}_{\rm usf}$ & $\gamma^{123}_{\rm usf}$ & $\tau^{110}_{\rm iss}$ & $\tau^{112}_{\rm iss}$ & $\tau^{123}_{\rm iss}$ \\
\midrule
BCC Hf  & 336.07 & 392.55 & 385.30 & 1.93 & 1.68 & 4.91 \\
BCC Ti  & 343.11 & 390.68 & 386.08 & 3.92 & 4.70 & 4.66 \\
BCC Zr  & 325.46 & 369.17 & 363.33 & 3.73 & 4.41 & 4.38 \\
BCC Mo  & 1458.05 & 1689.03 & 1657.93 & 33.10 & 58.40 & 19.30 \\
BCC Nb  & 604.87 & 697.23 & 684.71 & 14.10 & 24.50 & 8.10 \\
BCC Ta  & 751.11 & 868.30 & 852.26 & 8.74 & 10.14 & 9.94 \\
\bottomrule 
\end{tabular}
\end{table}

Furthermore, the three lowest USFE and ISS values for all three slip planes consistently follow the same order, i.e., $\mathrm{NbTiZr} < \mathrm{HfNbTi} < \mathrm{TaTiZr}$. Extending to the five lowest values for both USFE and ISS adds HfTaTi and HfNbTaTiZr, although their precise rankings differ by slip plane. Notably, these five compositions each contain an HCP-element percentage exceeding 50\%, indicating that a high fraction of HCP elements can reduce both USFE and ISS in BCC RMPEAs. To further illustrate this point, the USFE and the ISS for BCC Hf, BCC Ti, BCC Zr and the other three conventional BCC metals on the three types of slip planes (\{110\}, \{112\} and \{123\} planes) are shown in \autoref{tbl:USFE_ISS}. Compared to the conventional BCC metals (Mo, Nb, Ta), the BCC phases of Hf, Ti, and Zr exhibit significantly lower USFEs and ISSs. Consequently, the fraction of HCP elements in the BCC RMPEA composition above 50\% significantly lowers both USFE and ISS.

Apart from those RMPEAs that achieve the highest and lowest values, the ISS values do not follow the same ranking as the USFE values (\autoref{fig:3}(b) and \autoref{tbl:mech_sum}). These observations imply that the effect of composition on ISS is more intricate than on USFE and may not be dominated by a single element. One reason for this difference is because the ISS involves elastic deformation of the entire material volume, whereas the USFE pertains to breaking atomic bonds that span across a single crystallographic plane in the crystal.

The experimental measurement of stacking fault energy (SFE) in BCC alloys, particularly in RMPEAs, presents significant challenges due to the intrinsic nature of BCC crystal structures. As noted in traditional metallurgical theory \citep{hirth1982theory}, BCC materials typically exhibit high SFE values, resulting in extremely narrow separation between dislocation partials (often less than 1~nm) that effectively merge into complete dislocations. This fundamental characteristic renders direct observation of extended stacking faults via high-resolution transmission electron microscopy (HRTEM) nearly impossible in conventional BCC systems. Our computational approach employing GSFE curves follows established methodology in the field \citep{vitek2011pms,wang2021cms}, where atomistic simulations serve as the primary tool for understanding shear resistance mechanisms. The USFE and ISS values presented in this study should therefore be interpreted as theoretical proxies for slip resistance rather than experimentally observable quantities. Despite advancements in the measurement of the dislocation density in BCC metals using TEM techniques \citep{meng2021mc}, the direct visualization of extended stacking faults in BCC-based RMPEAs remains technically unfeasible due to the negligible separation between partial dislocations. This limitation underscores the necessity of computational approaches in investigating dislocation mechanisms in these novel alloy systems.

\subsubsection{Zener ratio}
%%%Fig4
\begin{figure}[htb]
\centering
\includegraphics[scale=0.4,clip]{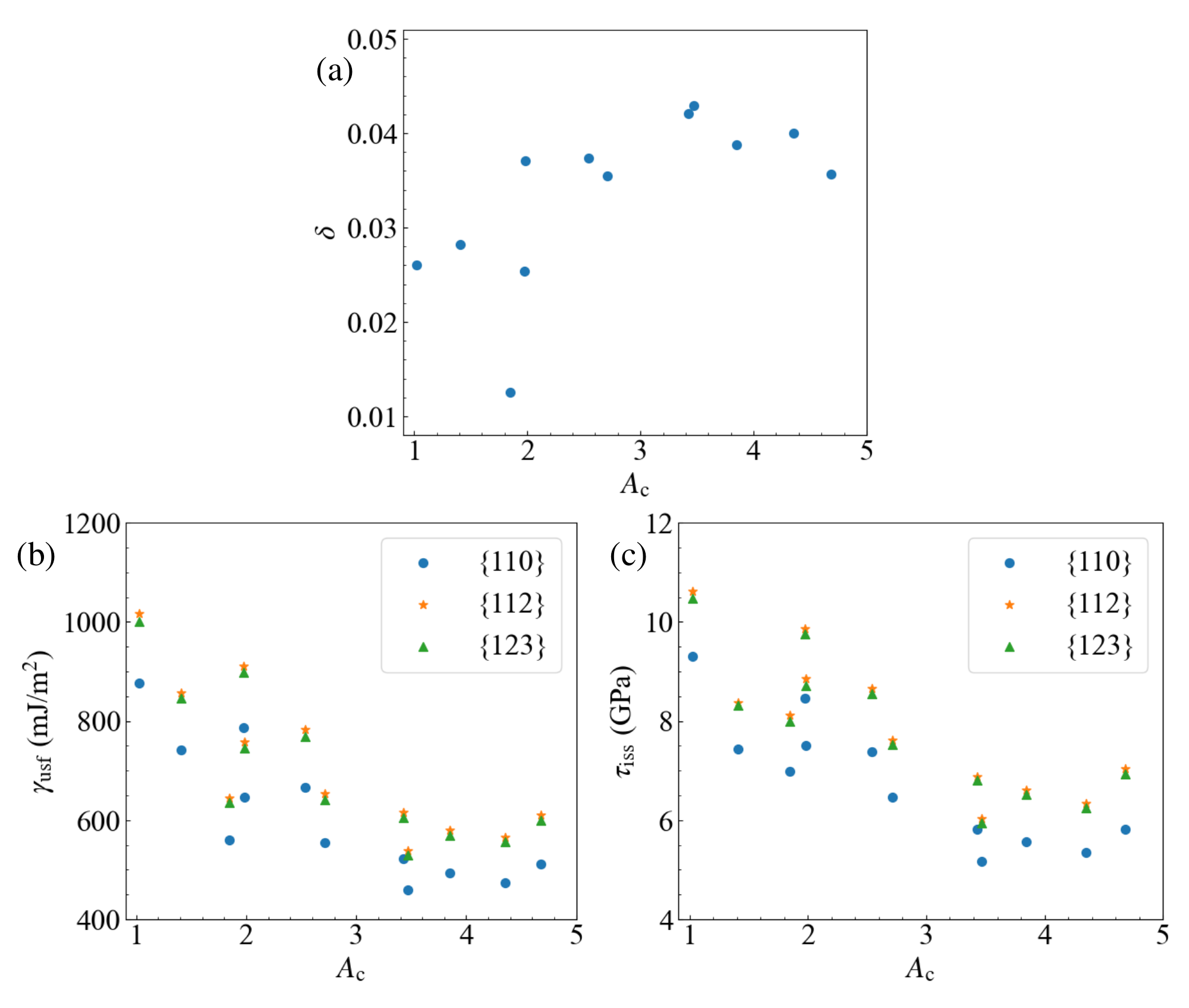}
\caption{The (a) LD coefficient ($\delta$), (b) USFE ($\gamma_{\mathrm{usf}}$) and (c) ISS ($\tau_{\mathrm{iss}}$) as a function of Zener ratio ($A_{\rm c}$) for 12 RMPEAs. $\{110\}$, $\{112\}$, and $\{123\}$ represent three slip planes.}
\label{fig:4}
\end{figure}

\autoref{tbl:mech_sum} presents the average elastic constants ($C_{11}$, $C_{12}$ and $C_{44}$) and Zener ratio ($A_{\rm c}$) for 12 RMPEAs. The Zener ratio, defined as $2C_{44}/(C_{11}-C_{12})$, is commonly used to assess the degree of elastic anisotropy in cubic crystals. A value of 1 indicates an isotropic material, whereas greater deviations below or above unity signify increasing anisotropy. For the 12 RMPEAs studied, the Zener ratio ranges from 1 to 5.

Considering the importance of LD ($\delta$), USFE ($\gamma_{\mathrm{usf}}$), and ISS ($\tau_{\mathrm{iss}}$) in dislocation glide \citep{xu2021acta,romero2022ijp}, we study in \autoref{fig:4} their relationships with the Zener ratio. As shown in \autoref{fig:4}(a), the Zener ratio and LD coefficient exhibit a positive correlation when the Zener ratio lies between 1 and 3.5, beyond which further increases in Zener ratio are uncorrelated with LD. In contrast, USFE and ISS decrease as the Zener ratio increases from 1 to 3.5 and plateaus for higher values, implying that elevated elastic anisotropy may reduce the resistance to dislocation glide.

\begin{figure}[htb]
\centering
\includegraphics[scale=0.5,clip]{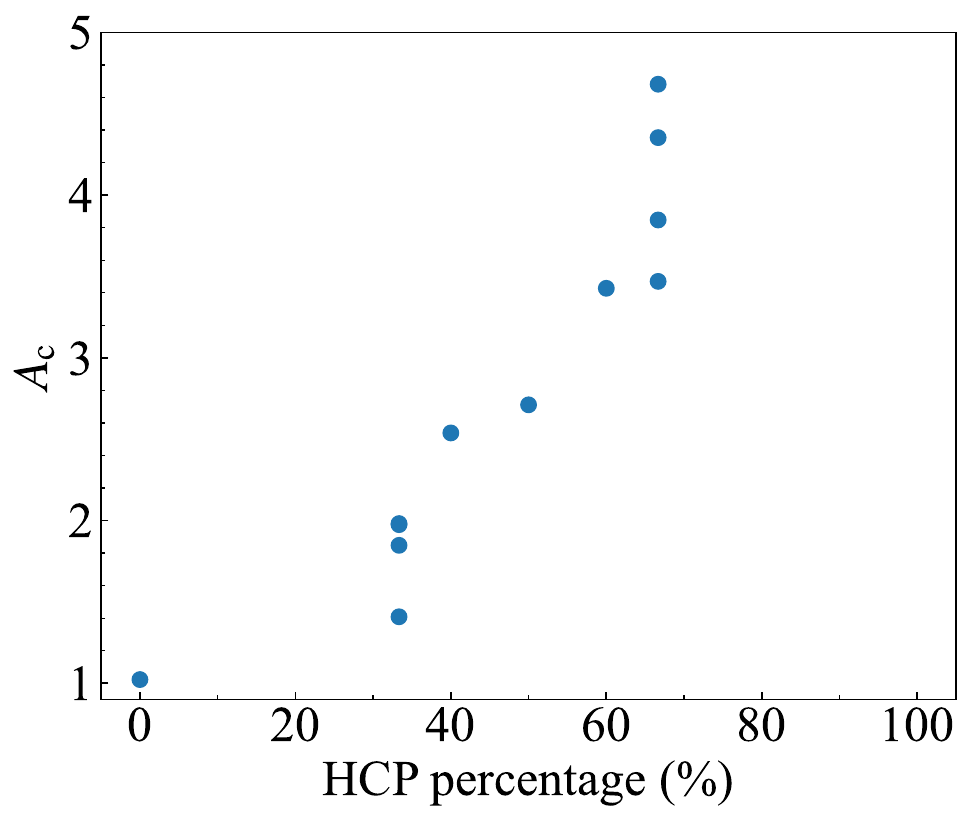}
\caption{The Zener ratio ($A_{\rm c}$) as a function of HCP element percentage for 12 RMPEAs. The data points for HfNbTa (33.33, 1.98) and MoTaTi (33.33, 1.97) nearly overlap.}
\label{fig:HCP_Zener}
\end{figure}

In addition, the Zener ratio, \(A_{\text{c}}\), as a function of the HCP-element percentage, is also shown in \autoref{fig:HCP_Zener}. It can be observed that the Zener ratio increases rapidly with rising HCP-element percentage, indicating that the degree of elastic anisotropy in the cubic crystals grows accordingly. Based on the results in \autoref{fig:4}(b--c), both the USFE and the ISS decrease as the Zener ratio increases. Therefore, the increase in HCP-element percentage leads to a reduction in USFE and ISS, which is also attributed to the enhancement of elastic anisotropy.

\subsection{LSRs of RMPEAs}
\subsubsection{LSR analysis in 12 RMPEAs}
%%%Fig5
\begin{figure}[htp]
\centering
\includegraphics[scale=0.7,clip]{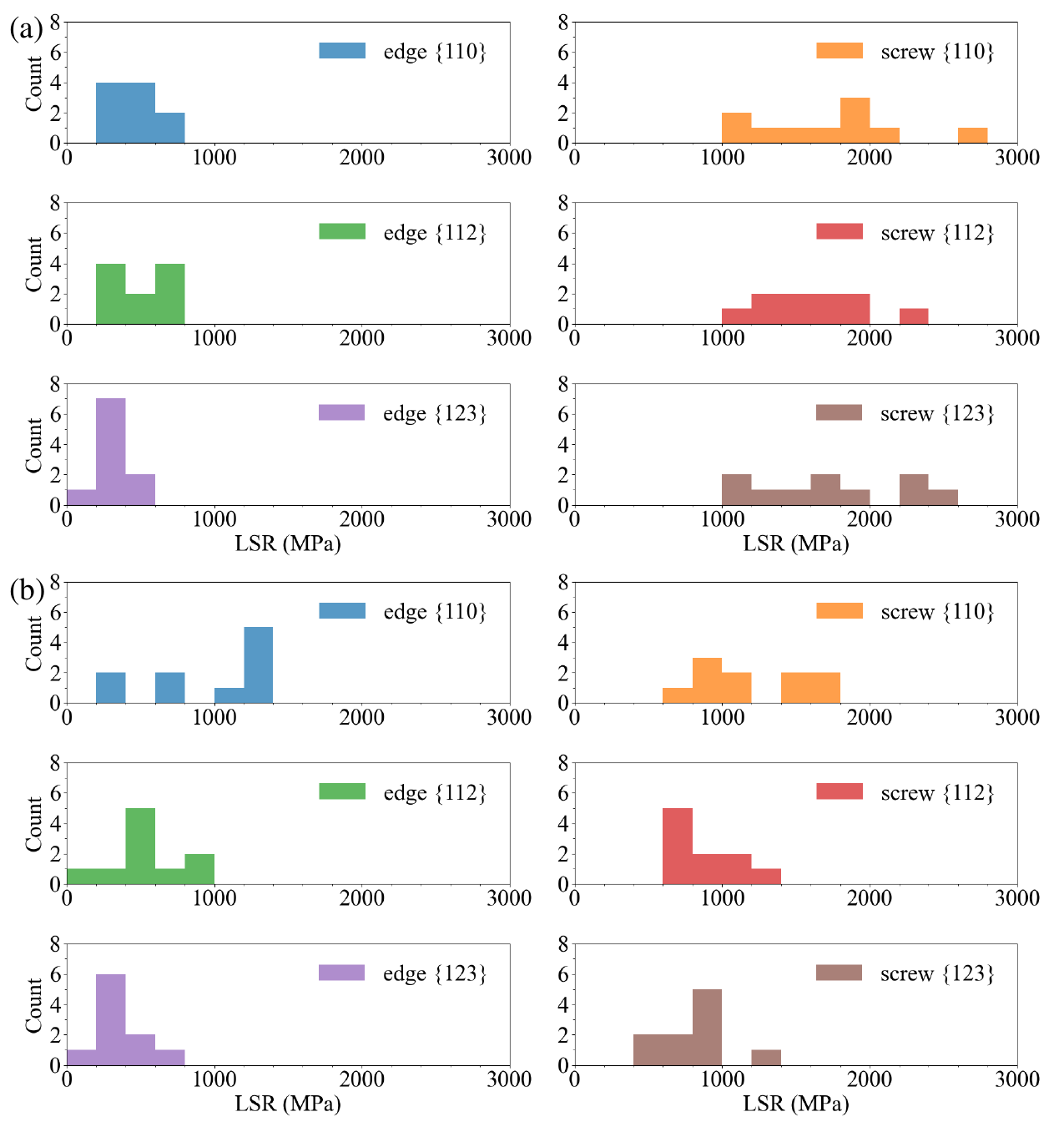}
\caption{Statistical distribution of LSR values for (a) NbTaTi and (b) NbTiZr RMPEAs. For each alloy, results are shown for 10 independent atomic configurations, with distributions plotted separately for edge and screw dislocations on the $\{110\}$, $\{112\}$, and $\{123\}$ slip planes.}
\label{fig:5}
\end{figure}

As outlined in \autoref{sec:methodology}, we calculate the LSR values for both edge and screw dislocations on the $\{110\}$, $\{112\}$, and $\{123\}$ slip planes across 10 independent atomic configurations for each RMPEA composition. The distribution of these LSR values varies significantly among the different alloy compositions. \autoref{fig:5}(a) and (b) display the statistical distributions of LSR values for two representative cases: the NbTaTi alloy, which has the smallest LD coefficient ($\delta = 0.0126$) among the 12 studied RMPEAs, and the NbTiZr alloy, which has the largest LD coefficient ($\delta = 0.0429$). A direct comparison between them reveals different statistical scatter. For the low-distortion NbTaTi alloy, the LSR distributions are relatively narrow and concentrated for edge dislocations but are more dispersed for screw dislocations across all three glide planes. Conversely, in the high-distortion NbTiZr alloy, this trend is altered: the LSR distributions become relatively broader and more dispersed for edge dislocations, while appearing more concentrated for screw dislocations. Dislocation motion in RMPEAs is influenced by the composition and configuration of atoms in the local atomic environments surrounding the dislocations and the associated local energy barriers \citep{xu2021acta, romero2022ijp}. Since LD directly affects these energy barriers, we hypothesize that the statistical dispersion in the LSR distribution is closely related to LD.

%%%Table3
\begin{table*}[]
\rotatebox{90}{
\begin{minipage}{\textheight} % Keep the table aligned and inside page margins
\caption{The LSR values (MPa) of edge and screw dislocations on the $\{110\}$, $\{112\}$, and $\{123\}$ planes for 12 RMPEAs from MS calculations. The Peierls stresses (MPa) of BCC Mo, Nb and Ta from MS calculations are also shown for comparison.}
\label{tbl:lsr_sum}
\centering
\resizebox{\textwidth}{!}{
\begin{tabular}{c|ccccccccccccc}
\hline
\diagbox{Composition}{Values}  & $\tau^{110}_{\mathrm{e}}$ & $\tau^{110}_{\mathrm{s}}$ & $\tau^{112}_{\mathrm{e}}$ & $\tau^{112}_{\mathrm{s}}$ & $\tau^{123}_{\mathrm{e}}$ & $\tau^{123}_{\mathrm{s}}$ & $\frac{\tau^{110}_{\mathrm{s}}}{\tau^{110}_{\mathrm{e}}}$ & $\frac{\tau^{112}_{\mathrm{s}}}{\tau^{112}_{\mathrm{e}}}$ & $\frac{\tau^{123}_{\mathrm{s}}}{\tau^{123}_{\mathrm{e}}}$ & $\frac{\tau^{110}_{\mathrm{e}}}{\tau^{112}_{\mathrm{e}}}$ & $\frac{\tau^{110}_{\mathrm{s}}}{\tau^{112}_{\mathrm{s}}}$ & $\frac{\tau^{110}_{\mathrm{e}}}{\tau^{123}_{\mathrm{e}}}$ & $\frac{\tau^{110}_{\mathrm{s}}}{\tau^{123}_{\mathrm{s}}}$\\
\hline
NbTaTi      & 443.03 & 1731.11 & 514.18 & 1616.89 & 340.94 & 1737.92 & 3.91 & 3.14 & 5.10 & 0.86 & 1.07 & 1.30 & 1.00 \\
MoNbTi      & 797.49 & 2016.97 & 1085.82 & 1634.26 & 795.84 & 1776.54 & 2.53 & 1.51 & 2.23 & 0.73 & 1.23 & 1.00 & 1.14 \\
HfNbTa      & 777.59 & 1394.83 & 708.49 & 1104.07 & 592.07 & 1032.13 & 1.79 & 1.56 & 1.74 & 1.10 & 1.26 & 1.31 & 1.35 \\
NbTiZr      & 956.74 & 1226.94 & 556.23 & 877.68 & 344.02 & 811.20 & 1.28 & 1.58 & 2.36 & 1.72 & 1.40 & 2.78 & 1.51 \\
HfNbTi      & 617.74 & 1021.31 & 590.82 & 994.67 & 413.33 & 674.14 & 1.65 & 1.68 & 1.63 & 1.05 & 1.03 & 1.49 & 1.51 \\
HfTaTi      & 912.06 & 1013.25 & 758.30 & 1012.44 & 549.66 & 740.18 & 1.18 & 1.34 & 1.35 & 1.13 & 1.00 & 1.56 & 1.37 \\
TaTiZr      & 1138.82 & 1414.79 & 732.28 & 1098.29 & 474.18 & 842.95 & 1.24 & 1.50 & 1.78 & 1.56 & 1.29 & 2.40 & 1.68 \\
MoTaTi      & 922.53 & 1975.66 & 1185.92 & 1686.38 & 736.58 & 1742.48 & 2.14 & 1.42 & 2.37 & 0.78 & 1.17 & 1.25 & 1.13 \\
MoNbTa      & 949.41 & 2320.54 & 1303.57 & 1645.52 & 798.93 & 1977.64 & 2.44 & 1.26 & 2.48 & 0.73 & 1.41 & 1.19 & 1.17 \\
HfNbTaTi    & 884.90 & 1155.84 & 637.74 & 1062.53 & 624.89 & 1220.25 & 1.31 & 1.67 & 1.95 & 1.39 & 1.09 & 1.42 & 0.95 \\
HfMoNbTaTi  & 1543.22 & 2017.60 & 878.73 & 1772.92 & 772.36 & 1395.88 & 1.31 & 2.02 & 1.81 & 1.76 & 1.14 & 2.00 & 1.45 \\
HfNbTaTiZr  & 923.91 & 1084.82 & 659.58 & 1021.67 & 458.90 & 782.21 & 1.17 & 1.55 & 1.70 & 1.40 & 1.06 & 2.01 & 1.39 \\
Mo  & 50.40 & 3611.46 & 459.9 & 2683.79 & 130.39 & 3052.18 & 71.66 & 5.84 & 23.41 & 0.11 & 1.35 & 0.39 & 1.18 \\
Nb  & 7.53 & 920.2 & 99.5 & 854.01 & 4.70 & 773.03 & 122.20 & 8.58 & 164.47 & 0.08 & 1.08 & 1.60 & 1.19 \\
Ta  & 13.13 & 3224.5 & 274.09 & 2024.29 & 29.86 & 2204.2 & 245.58 & 7.39 & 73.82 & 0.05 & 1.59 & 0.44 & 1.46 \\
\hline
\end{tabular}
}
\end{minipage}
}
\end{table*}

%%%Fig6
\begin{figure}[htp]
\centering
\includegraphics[scale=0.35,clip]{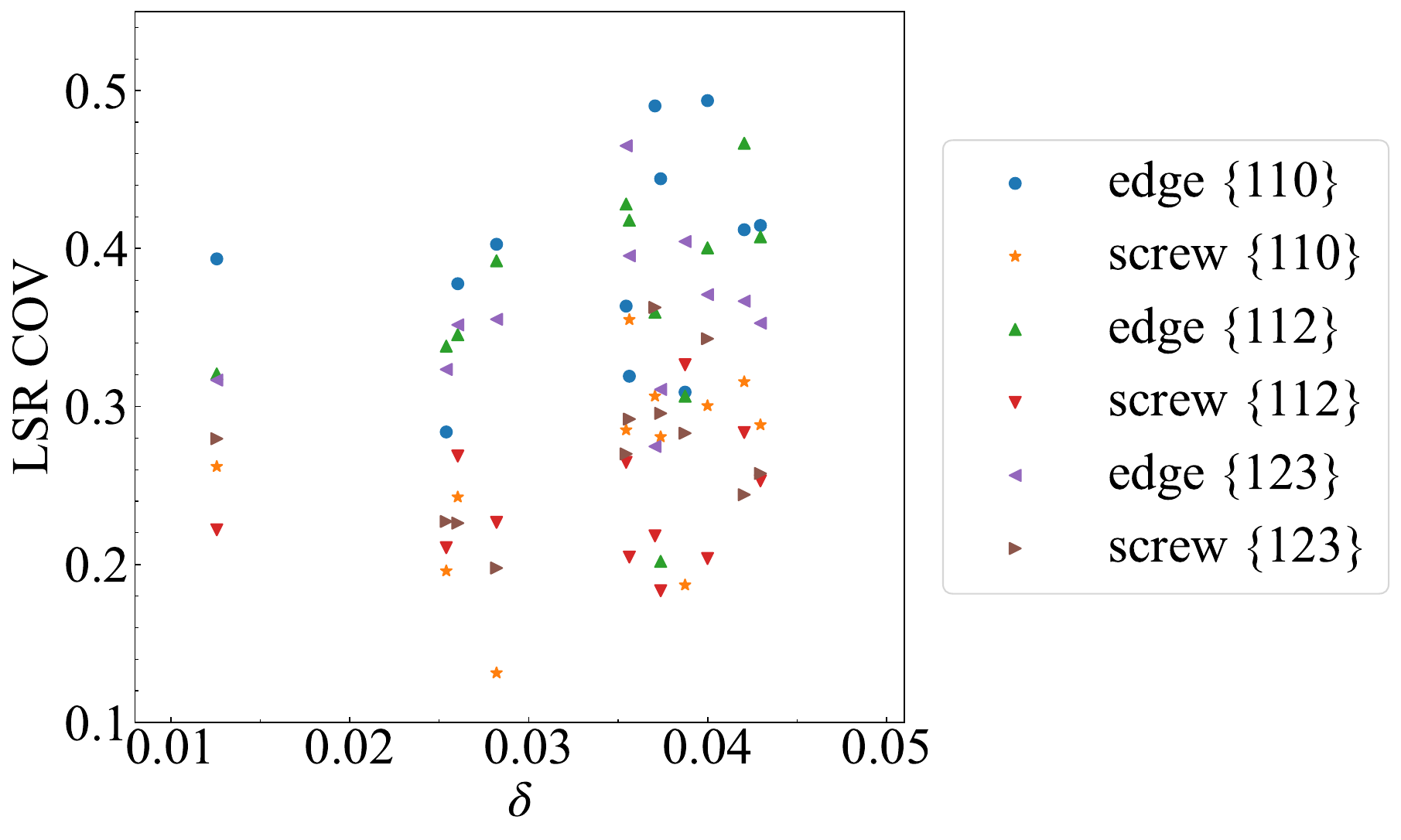}
\caption{The COVs of the LSR values on $\{110\}$, $\{112\}$ and $\{123\}$ slip planes in 12 RMPEAs as a function of the LD coefficient ($\delta$).}
\label{fig:6}
\end{figure}

To investigate this relationship, we plot the coefficient of variation (COV) of LSR values against the LD coefficient ($\delta$) in \autoref{fig:6} for both edge and screw dislocations across three slip planes in all 12 RMPEAs. The COV, defined as the ratio of the standard deviation to the mean value, is a normalized measure of dispersion in the LSR distribution, with a higher COV indicating greater variability. The results indicate that the COVs of LSR values remain relatively unchanged as $\delta$ increases from 0.01 to 0.35. However, beyond $\delta = 0.35$, the COVs increase significantly, suggesting that a very large LD coefficient leads to greater variation in LSR for all three slip planes and for both edge and screw dislocations alike.

%%%Fig7
\begin{figure}[htb]
\centering
\includegraphics[scale=0.55,clip]{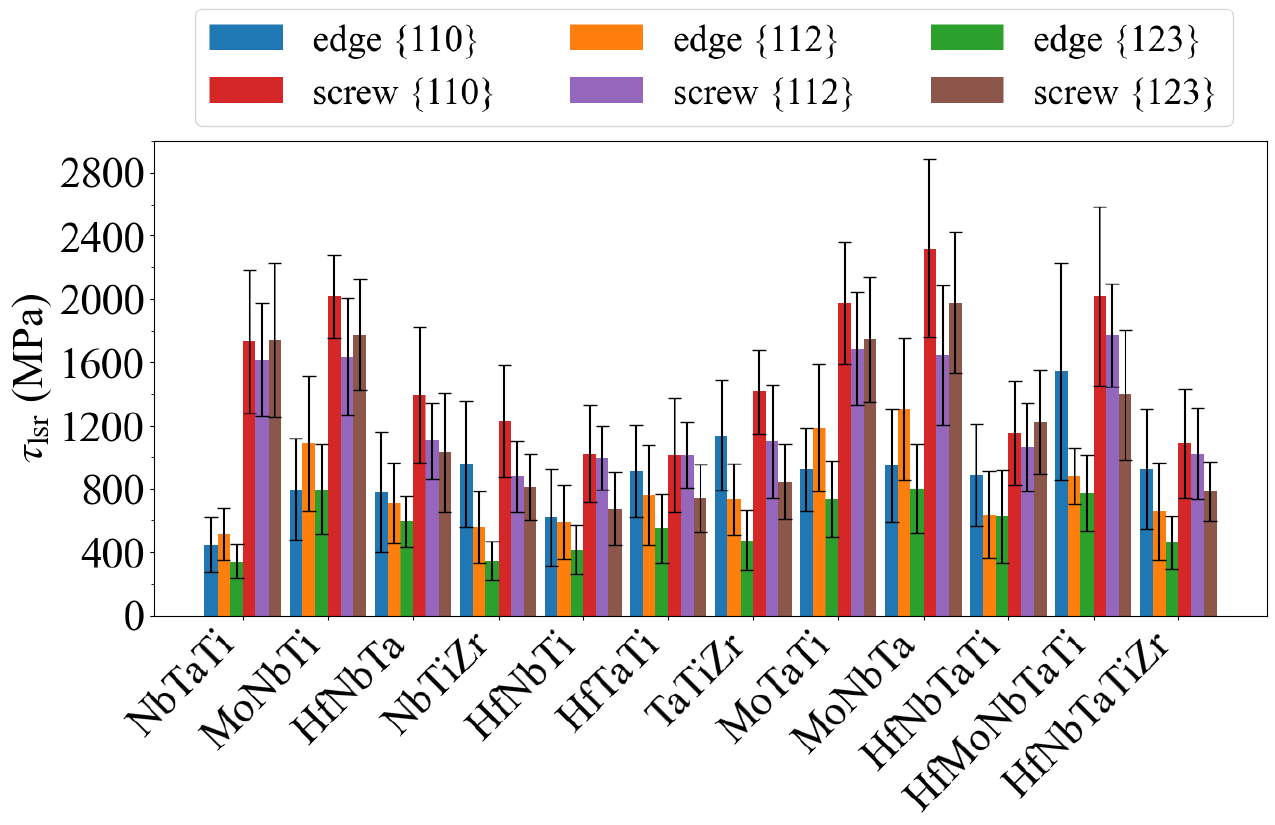}
\caption{The mean LSR values for $\{110\}$, $\{112\}$ and $\{123\}$ slip planes in 12 RMPEAs.}
\label{fig:7}
\end{figure}

The average LSR values over 10 samples for the three slip planes of all 12 RMPEA compositions are presented in \autoref{fig:7} and \autoref{tbl:lsr_sum}. For comparison, the Peierls stresses of edge and screw dislocations on the same three slip planes for BCC Mo, Nb and Ta are also shown in \autoref{tbl:lsr_sum}. Among all RMPEAs, NbTaTi exhibits the lowest LSR values for edge dislocations across all three slip planes, specifically 443.03 MPa, 514.18 MPa, and 340.94 MPa for the $\{110\}$, $\{112\}$ and $\{123\}$ slip planes, respectively. In contrast, the HfMoNbTaTi alloy shows the highest LSR value for edge dislocation on the \{110\} slip plane (1543.22 MPa), while the MoNbTa composition records the largest LSR values for edge dislocations on the \{112\} and \{123\} slip planes, with values of 1303.57 MPa and 798.93 MPa, respectively. For screw dislocations, the lowest LSR values for various slip planes are observed in different RMPEAs: HfTaTi for the \{110\} slip plane (1013.25 MPa), NbTiZr for the \{112\} slip plane (877.68 MPa), and HfNbTi for the \{123\} slip plane (674.14 MPa). Additionally, the MoNbTa alloy exhibits the highest LSR values for screw dislocations on both the \{110\} and \{123\} slip planes, with values of 2320.54 MPa and 1977.64 MPa, respectively. The HfMoNbTaTi alloy also shows the largest LSR value for screw dislocation on the \{112\} slip plane (1772.92 MPa). As shown in \autoref{tbl:lsr_sum}, compared to RMPEAs, BCC pure metals Mo, Nb, and Ta exhibit significantly lower LSR values for edge dislocations and higher LSR values for screw dislocations.

Overall, the NbTaTi composition leads to the weakest LSR for edge dislocations across the three slip planes, while the BCC elements-containing MoNbTa composition results in the strongest LSR for both edge and screw dislocations in MoNbTa and HfMoNbTaTi alloys. In contrast, the Hf, Ti or Zr-containing alloys with a high percentage of HCP elements (exceeding 50\%) exhibit low LSR values for screw dislocations on all three slip planes. Specifically, the lowest LSR values for screw dislocations on the \{110\}, \{112\}, and \{123\} slip planes are observed in HfTaTi, NbTiZr, and HfNbTi, respectively.

The validity and reliability of the LSR values obtained from the atomistic simulations indeed rest on how the local chemical environment, including solute segregation and atomic arrangement, affects the intrinsic resistance to dislocation glide. Crucially, the approach in this work does not rely on any a priori assumption about the additivity or interaction between a ``pure‑metal'' Peierls stress and solute effects. Instead, the LSR is computed directly from the atomic‑scale energy landscape of a realistic, chemically disordered structure in RMPEAs. In our simulations, a short (1-2 nm) dislocation segment is embedded in a fully random solid‑solution configuration, and the critical stress required to move that segment is calculated for many different local atomic environments. The resulting distribution of LSR values therefore inherently captures the spatially varying influence of solute atoms, including any local segregation, on the dislocation‑line energy barrier. Because we explicitly sample the local atomic configurations that a dislocation would encounter in a random RMPEA, our LSR values are a direct measure of the effective local lattice friction, without the need to assume a separate ``Peierls stress'' that is then modified by solute effects. In this way, our simulation‑based LSR provides a reliable, atomistically grounded descriptor of the intrinsic slip resistance in RMPEAs.

\subsubsection{LSR versus fundamental properties}
%%%Fig8
\begin{figure}[htb!]
\centering
\includegraphics[scale=0.48,clip]{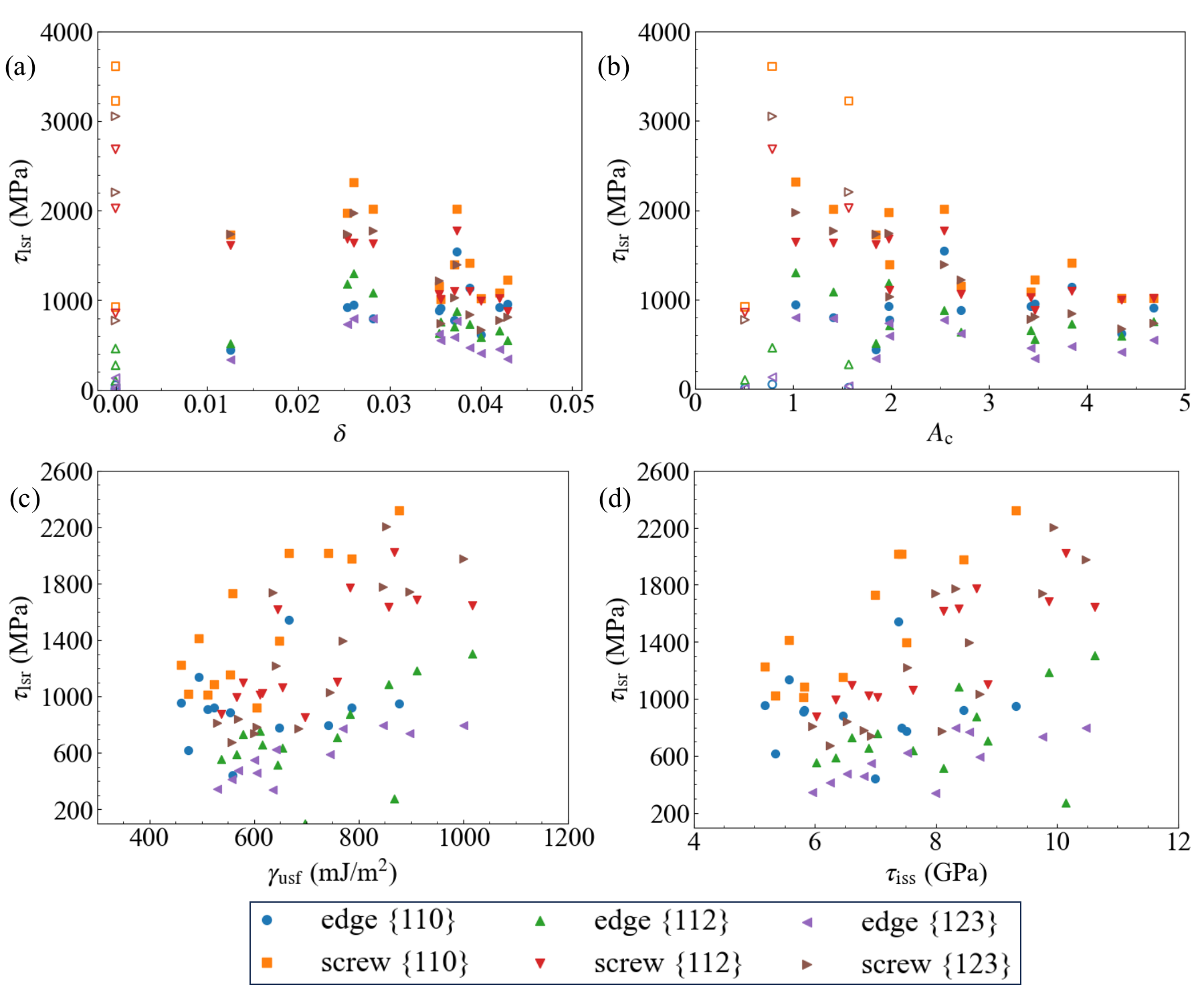}
\caption{The mean LSR on $\{110\}$, $\{112\}$ and $\{123\}$ slip planes in 12 RMPEAs as a function of (a) LD coefficient ($\delta$), (b) Zener ratio ($A_{\rm c}$), (c) USFE ($\gamma_{\mathrm{usf}}$), and (d) ISS ($\tau_{\mathrm{iss}}$), respectively. The open symbols in (a) and (b) denote data points for BCC pure metals (Mo, Nb, Ta) where the Peierls stress, rather than the LSR value, applies.}
\label{fig:8}
\end{figure}

Based on the analysis of the fundamental properties of RMPEAs in \autoref{sec:fundamental_properties}, the mean LSR values for all 12 RMPEA compositions are plotted against their LD coefficient ($\delta$), Zener ratio ($A_{\rm c}$), USFE ($\gamma_{\mathrm{usf}}$), and ISS ($\tau_{\mathrm{iss}}$) in \autoref{fig:8}. For comparison, data points for BCC pure metals (Mo, Nb, Ta) using Peierls stress instead of LSR value have been added to (a) and (b). As demonstrated in \autoref{fig:8}(a), while no clear linear correlation exists between LD coefficient ($\delta$) and LSR values for edge dislocations in RMPEAs, the LSR values for screw dislocations exhibit a general decreasing trend with increasing $\delta$. The inclusion of data for BCC pure metals ($\delta = 0$) reinforces this latter trend. Similarly, \autoref{fig:8}(b) shows that the LSR values across all three slip planes exhibit a decreasing trend with increasing Zener ratio ($A_{\rm c}$) in RMPEAs, although the rates of decrease vary for different dislocation types and slip planes. With the inclusion of BCC pure metal data, this trend persists only for screw dislocations and not for edge dislocations. This finding indicates that the mobility mechanisms of screw versus edge dislocations differ more significantly in BCC pure metals than they do in RMPEAs. Furthermore, as elucidated in \autoref{fig:4}(b) and (c), the USFE ($\gamma_{\mathrm{usf}}$) and ISS ($\tau_{\mathrm{iss}}$) are negatively correlated with the Zener ratio across all three slip planes in RMPEAs. Consequently, \autoref{fig:8}(c) and (d) reveal that the LSR value is positively correlated with both USFE and ISS. This shows that higher USFE and ISS values are associated with increased LSR, highlighting the multifaceted factors that influence LSR in RMPEAs beyond $\delta$ and $A_{\rm c}$.

Based on the above results, the observation in \autoref{fig:7} that increasing the fraction of HCP elements in the RMPEA composition above 50\% significantly reduces the LSR values for screw dislocations on all three slip planes can be analyzed from the following perspectives. First, the reduction in the USFE ($\gamma_{\text{usf}}$) and the ISS ($\tau_{\text{iss}}$) due to the incorporation of HCP elements directly decreases the energy barrier for the motion of screw dislocations in BCC RMPEAs, which is consistent with the results shown in \autoref{fig:8}(c) and (d). The lattice resistance to glide, quantified by the LSR, is intrinsically linked to the energy landscape of the slip plane. Within the framework of thermally activated glide, a lower $\gamma_{\text{usf}}$ reduces the energy penalty for dislocation glide \citep{romero2022ijp}, which correlates with a lower Peierls barrier that the screw dislocation must overcome. Consequently, the critical stress to move a dislocation at 0 K is reduced, leading to a lower LSR. Similarly, a lower $\tau_{\text{iss}}$, representing the maximum sustainable lattice stress, implies a reduced intrinsic strength of the perfect lattice against shear \citep{romero2022ijp}, thereby lowering the critical stress for dislocation motion. Second, \autoref{fig:HCP_Zener} shows that as the HCP-element percentage rises, \(A_{\text{c}}\) increases sharply, reflecting a corresponding growth in the elastic anisotropy of the cubic crystals. As shown in \autoref{fig:8}(b), the LSR values across all three slip planes exhibit a decreasing trend with increasing Zener ratio ($A_{\rm c}$). Consequently, a higher HCP-element percentage reduces the LSR across all considered slip planes, an effect that can likewise be attributed to the increased elastic anisotropy. Together, these mechanistic links explain the characteristically low LSR values for screw dislocations observed in Hf-, Ti-, or Zr-rich RMPEAs with high HCP-element fractions.

\autoref{fig:8} demonstrates LD, Zener ratio, USFE and ISS play a significant role in determining the LSR value. These findings also underscore the complexity of the mechanisms governing LSR in RMPEAs, suggesting that multiple factors must be considered to fully understand and predict dislocation behavior in these alloys.

\subsubsection{Plasticity anisotropy}
%%%Fig9
\begin{figure}[htp]
\centering
\includegraphics[scale=0.42,clip]{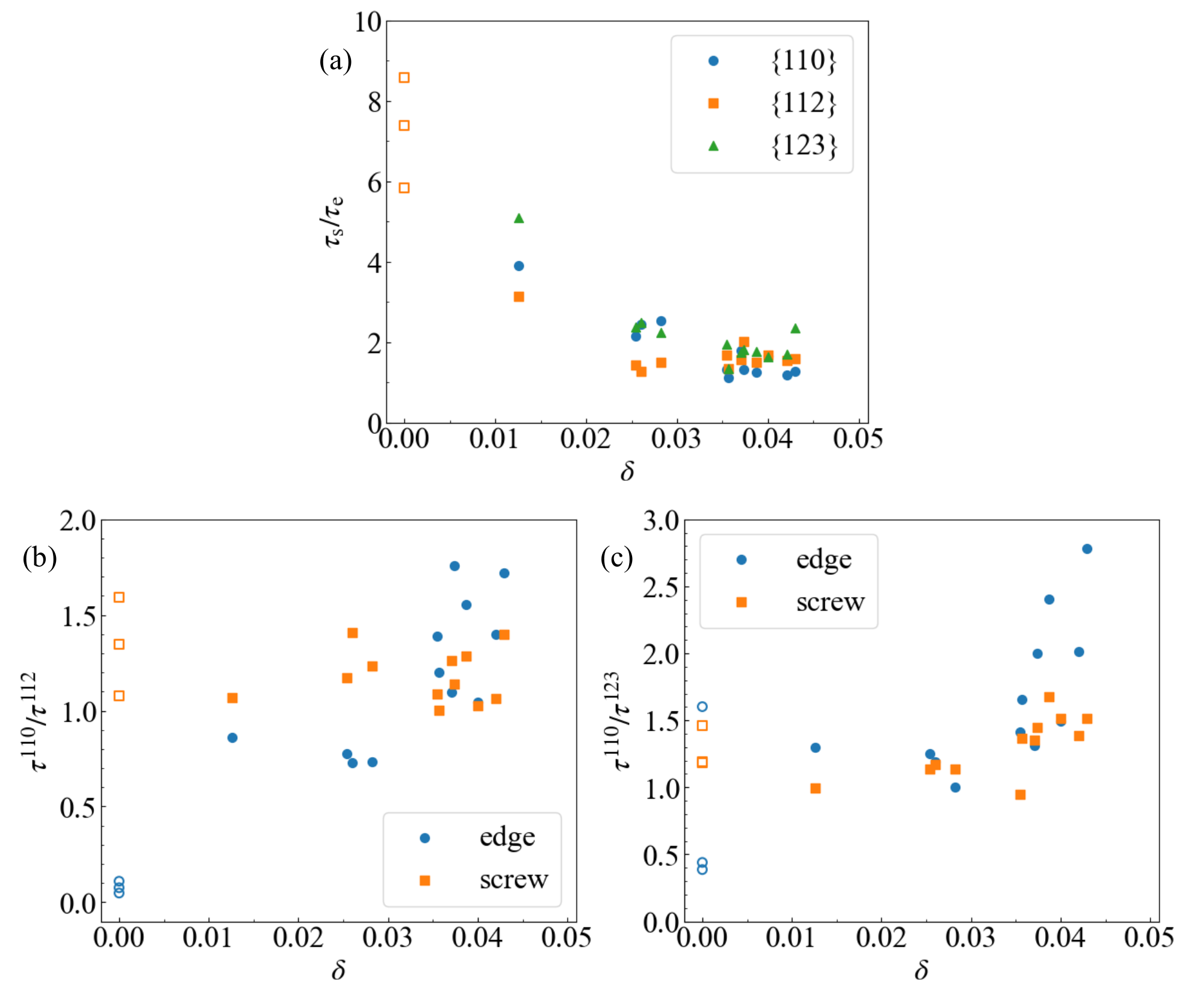}
\caption{(a) Screw-to-edge LSR ratios across three glide planes, (b) \{110\}-to-\{112\} LSR ratios and (b) \{110\}-to-\{123\} LSR ratios for edge and screw dislocations, plotted as a function of the LD coefficient ($\delta$) in 12 RMPEAs. The open symbols denote data points for BCC pure metals (Mo, Nb, Ta) where the Peierls stress, rather than the LSR value, applies. Given that the screw-to-edge ratios of Peierls stress on \{110\} and \{123\} slip planes are extremely high for BCC pure metals (\autoref{tbl:lsr_sum}), the corresponding data points are not shown in (a).}
\label{fig:9}
\end{figure}

In addition to LSR, we examine plasticity anisotropy, which is characterized by the ratios between different types of LSR values. As shown in \autoref{fig:9}(a), the screw-to-edge LSR ratios across various slip planes decrease with increasing LD coefficient ($\delta$) in RMPEAs. The inclusion of Peierls stress data for BCC pure metals ($\delta = 0$) strengthens this trend considerably. Given that their screw-to-edge ratios of Peierls stress on \{110\} and \{123\} slip planes are extremely high (\autoref{tbl:lsr_sum}), the corresponding data points are not shown in (a) for a better visualization of this trend on a clear scale.} Specifically, as $\delta$ rises, the screw-to-edge ratios of RMPEAs decline from approximately 3-5 to a range between 1 and 2, indicating a reduction in plasticity anisotropy between different dislocation types with the more severe LD. A prior work reported a similar phenomenon for the LSR on \{110\} planes in some RMPEAs \citep{wang2024npjcm}. Here, we confirm it for higher-order planes. The reduction in the screw-to-edge LSR ratio under severe LD originates from the fundamentally different ways edge and screw dislocation cores interact with a chemically disordered lattice. The core of a screw dislocation in BCC RMPEAs has a non-planar, three-dimensional structure, making it highly sensitive to the local shear stress field. In an RMPEA, the random atomic distribution creates a spatially fluctuating stress field. This heterogeneity can apply local torques that provide alternative, lower-energy paths for screw segments to advance, thereby reducing their average critical stress. In contrast, the planar core of an edge dislocation is less effective at utilizing these local fluctuations to find easier paths. Therefore, as LD increases, the resistance of screw dislocations decreases more significantly relative to that of edge dislocations, leading to a lower screw-to-edge LSR ratio.

As another measure of anisotropy, \autoref{fig:9}(b) presents the \{110\}-to-\{112\} LSR ratios for both edge and screw dislocations. In RMPEAs, the \{110\}-to-\{112\} ratio for edge dislocation increases from around 0.7 to 1.8 with increasing the LD coefficient, whereas the ratio for screw dislocation fluctuates within the range of 1.0 to 1.5 as the LD coefficient increases. These two trends also hold for pure BCC metals ($\delta = 0$). This suggests that, under low LD conditions, the motions of edge dislocations are more easily activated on the \{110\} slip plane than on the \{112\} slip plane. However, as LD becomes more severe, this trend reverses, favoring the \{112\} slip plane. This reversal can be understood by considering the intrinsic slip resistance hierarchy in BCC crystals and how it is modulated by LD. In a perfect BCC lattice ($\delta = 0$), the Peierls stress for edge dislocations (with $\mathbf{b}=a/2\langle111\rangle$) is lower on \{110\} plane than \{112\} plane due to its larger interplanar spacing ($d_{110}=a/\sqrt{2}$ versus $d_{112}=a/\sqrt{6}$), higher planar atomic density, and smoother atomic topography along $\langle111\rangle$ directions. These factors lead to a shallower energy landscape on \{110\} plane, facilitating glide. However, with increasing LD (quantified by $\delta$), random atomic-size mismatch introduces local fluctuations in interatomic distances and bond strengths. The $\{112\}$ planes, with their lower symmetry and more variable interplanar spacing, are more sensitive to such perturbations. LD can create localized ``soft'' channels where the interplanar spacing is effectively increased or the shear resistance is reduced, thereby lowering the local LSR on $\{112\}$ plane relative to $\{110\}$ plane. Consequently, at high $\delta$, the intrinsic advantage of $\{110\}$ plane is overcome by disorder-induced softening on $\{112\}$ plane, reversing the LSR ratio. In contrast, screw dislocations consistently exhibit slightly easier activation on the \{112\} slip plane compared to the \{110\} slip plane, regardless of the LD coefficient. Additionally, the \{110\}-to-\{123\} LSR ratios in RMPEAs fluctuate between around 1.0 and 1.4 for both edge and screw dislocations as the LD coefficient increases to around 0.03 (\autoref{fig:9}(c)). Beyond this threshold, the ratios rapidly increase to around 2.8 for edge dislocations and to around 1.6 for screw dislocations with further increases in $\delta$. Generally, BCC pure metal data follows these trends. This indicates that when LD is not large in RMPEAs, dislocation motion is slightly easier to activate on the \{123\} slip plane than on the \{110\} slip plane for both edge and screw dislocations. This preference becomes markedly stronger for edge dislocations as LD becomes much more severe. The enhancement of the \{110\}-to-\{112\} and \{110\}-to-\{123\} LSR ratios with increasing LD can be understood by considering the atomic packing on different crystallographic planes. High-index planes such as \{112\} and \{123\} possess a lower in-plane atomic density (a more ``open'' structure) compared to the denser \{110\} plane, allowing dislocations to ``find'' easier paths through distorted lattices. Thus, under high LD, slip on these planes becomes relatively easier compared to \{110\} plane.

These findings suggest that the LD coefficient significantly influences plasticity anisotropy in RMPEAs. Specifically, severe LD leads to a greater number of dislocations being activated on high-index slip planes, particularly for edge dislocations. This highlights the complex interplay between LD and plasticity anisotropy, further underscoring the multifaceted factors that govern dislocation behavior in these alloys.

\section{Quantitative analysis}

The above results and discussion suggest that the LSR is qualitatively linked to the fundamental properties of RMPEAs, as well as their alloy compositions. To further explore this relationship, we first conduct a quantitative analysis using ML methods. Additionally, we propose a thermally activated, dislocation-based model to predict the tensile yield strength quantitatively, leveraging LSR values across the three most common slip planes: \{110\}, \{112\}, and \{123\}.

\subsection{Machine learning analysis}

\begin{figure}[htb!]
\centering
\includegraphics[scale=0.4,clip]{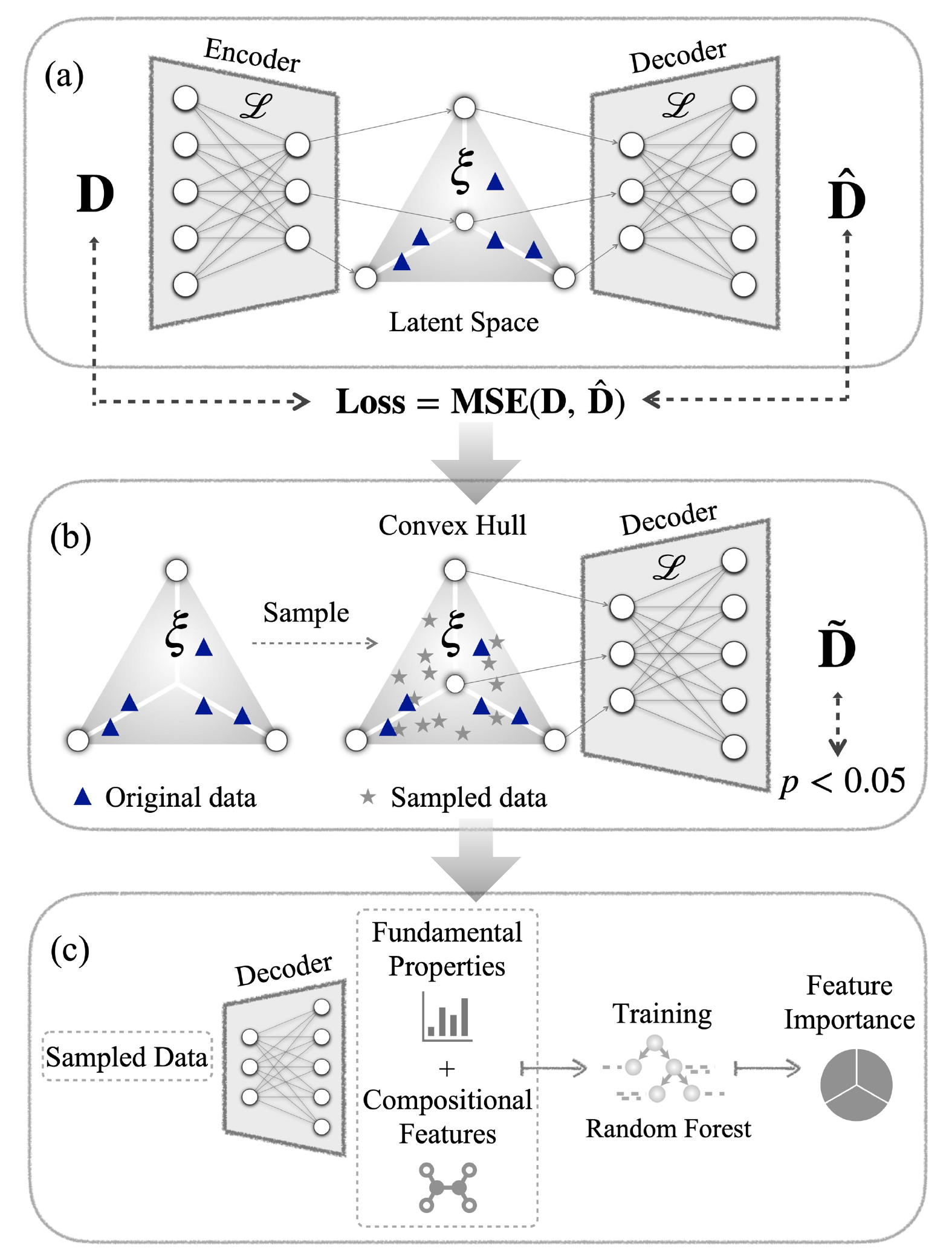}
\caption{Schematic of the ML analysis framework: (a) Autoencoder model architecture, (b) Data augmentation workflow, and (c) Random forest-based feature importance analysis.}
\label{fig:10}
\end{figure}

We train an autoencoder to reconstruct the input data and use the decoder to augment our limited dataset of material parameters (e.g., $C_{ij}$, $E_\text{coh}$, $a_0$, etc) for further analysis. Based on this augmented dataset, we then apply a random forest model to uncover the relationships between LSR and the intrinsic material properties and compositions of RMPEAs. The random forest algorithm is selected for the statistical and quantitative analysis in this study due to several key advantages that align with the specific characteristics of our dataset and research objectives. First, random forest algorithm naturally handles datasets with mixed feature types, seamlessly integrating continuous material properties (e.g., $C_{44}$, $\delta$) and categorical compositional data. Second, it provides robust feature importance rankings through metrics such as the mean decrease in accuracy, which is critical for identifying the dominant physical parameters governing LSR. Furthermore, the ensemble nature of random forest algorithm, which aggregates predictions from numerous decision trees, makes it inherently resistant to overfitting, a particularly valuable trait given the relatively limited size of our atomistically generated dataset compared to the feature space. The ML analysis framework is shown in \autoref{fig:10}.

%%%Fig10

\subsubsection{Autoencoder model training}
As shown in \autoref{fig:10}(a), we briefly outline the autoencoder training procedure.
Before training, we augment the existing dataset by performing 25 resampling iterations, drawing 10 samples per iteration.
All values are then rescaled to the order of $10^2$ before analysis.
Subsequently, we train the autoencoder on the preprocessed data.
The objective is to reconstruct the input data.
The training procedure for the autoencoder can be succinctly written as:
\begin{equation}
\begin{aligned}
\hat{\mathbf{D}} =\left(\underbrace{\mathcal{L}^{\mathbb{N}}\circ\mathcal{L}^{\mathbb{N}/2}\circ\mathcal{L}^3}_\text{Decoder}\circ\overbrace{\boldsymbol{\xi}}^\text{Eigenspace}\circ\underbrace{\mathcal{L}^3\circ\mathcal{L}^{\mathbb{N}/2}\circ\mathcal{L}^\mathbb{N}}_\text{Encoder}\right) \mathbf{D},
\end{aligned}
\label{equ:1}
\end{equation}
where we can denote $\mathbf{D}\equiv \left(\left[\delta, a_0, C_{ij}, E_{\rm coh}, \bm{\gamma}_{\rm usf}^{i}, \bm{\tau}_{\rm iss}^{i}\right],\ \left[\bm{\tau}_{\rm lsr}^{i}\right]\right)$ as the overall data, then \\$\hat{\mathbf{D}}\equiv \left(\left[\hat{\delta}, \hat{a}_0, \hat{C}_{ij}, \hat{E}_{\rm coh}, \hat{\bm{\gamma}}_{\rm usf}^{i}, \hat{\bm{\tau}}_{\rm iss}^{i}\right],\ \left[\hat{\bm{\tau}}_{\rm lsr}^{i}\right]\right)$ is the reconstructed data. 
$\mathcal{L}$ denotes the layers in the encoder and decoder. $\bm{\xi}$ is the latent space layer, where we choose to pick $\bm{\xi}\in\mathbb{R}^3$.
Additionally, we choose $\mathbb{N}=256$ as the dimension for both the encoder and the decoder.
The autoencoder compresses the material parameters into a latent representation, and the decoder decompresses this representation to reconstruct the original data.
The autoencoder is trained for $10^5$ epochs.
The trained autoencoder can capture the material parameters and regenerate new data with high accuracy, achieving a coefficient of determination ($R^2$) of 0.99 between $\mathbf{D}$ and $\hat{\mathbf{D}}$. This indicates that three latent dimensions are sufficient to capture the overall data structure.

\subsubsection{Data augmentation}
Once the autoencoder model training is completed, we hypothesize that the latent‑space representations lie within a convex hull.
By sampling the latent space, we can generate additional data sets for data augmentation, as shown in \autoref{fig:10}(b). The parameters taken for ML statistical analysis includes 12 fundamental properties ($\delta$, $a_0$, $C_{11}$, $C_{12}$, $C_{44}$, $E_{\mathrm{coh}}$, $\gamma^{110}_{\mathrm{usf}}$, $\gamma^{112}_{\mathrm{usf}}$, $\gamma^{123}_{\mathrm{usf}}$, $\tau^{110}_{\mathrm{iss}}$, $\tau^{112}_{\mathrm{iss}}$, and $\tau^{123}_{\mathrm{iss}}$) as well as 6 composition-dependent features (Hf, Ti, Zr, Nb, Mo, and Ta).
We randomly sample 1000 points from this convex hull and use the trained decoder to generate new data sets, denoted as $\tilde{\mathbf{D}}$.
Then, we filter these generated sets, retaining only those with statistically significant results ($p < 0.05$), where the $p$-value represents the probability of observing data at least as extreme as the results under the null hypothesis.
The data correlations of different material parameters are presented in Section~\ref{sec:data_corr}.

\subsubsection{Data correlations}
\label{sec:data_corr}
To incorporate the compositional information of an RMPEA into the feature space consisting of its fundamental properties, the alloy composition is decomposed into its constituent elements. Then, the presence of an element in the alloy is represented as a weighted fraction $(F(E_i))$, based on its atomic number $Z(E_i)$ that is normalized by the sum of the atomic numbers of all elements in the alloy:
\begin{equation}
F(E_i) = \frac{Z(E_i)/n}{\Sigma_{j=1}^n (Z(E_j)/n)},
\label{equ:2}
\end{equation}
where $n$ denotes the number of elements in the alloy, and $i \in \{1, \dots, n\}$.

The feature space for this study includes twelve fundamental properties ($\delta$, $a_0$, $C_{11}$, $C_{12}$, $C_{44}$, $E_{\mathrm{coh}}$, $\gamma^{110}_{\mathrm{usf}}$, $\gamma^{112}_{\mathrm{usf}}$, $\gamma^{123}_{\mathrm{usf}}$, $\tau^{110}_{\mathrm{iss}}$, $\tau^{112}_{\mathrm{iss}}$, and $\tau^{123}_{\mathrm{iss}}$) of RMPEAs along with six composition-dependent features of the alloy (Ti, Zr, Nb, Mo, Hf, and Ta). As shown in \autoref{fig:2}, the error bars included in the GSFE curves show that the standard deviations are negligible, averaging less than 1\% of the corresponding GSFE values. Given the minimal spread observed, this variability is not considered a significant feature for the ML model, and therefore the dispersion of the GSFE data is not included as an input feature here.
For alloys containing fewer than six elements, the compositional features are generated by assigning a zero value to the weighted fraction of any absent element. 
This enriched feature space is then employed for training, with the augmented dataset targeting six types of edge/screw dislocations on three glide planes.

This ML analysis focuses on feature importance to identify the fundamental properties or alloying elements that exert a significant influence on the LSR values of edge and screw dislocations across the three glide planes. 
As shown in \autoref{fig:10}(c), the random forest algorithm is utilized for this purpose.
Prior to conducting the feature importance analysis, each target variable is trained using the feature space to evaluate prediction confidence (quantified by the $R^2$ score) and the associated uncertainties.

%%%Fig11
\begin{figure}[htb!]
\centering
\includegraphics[scale=0.35,clip]{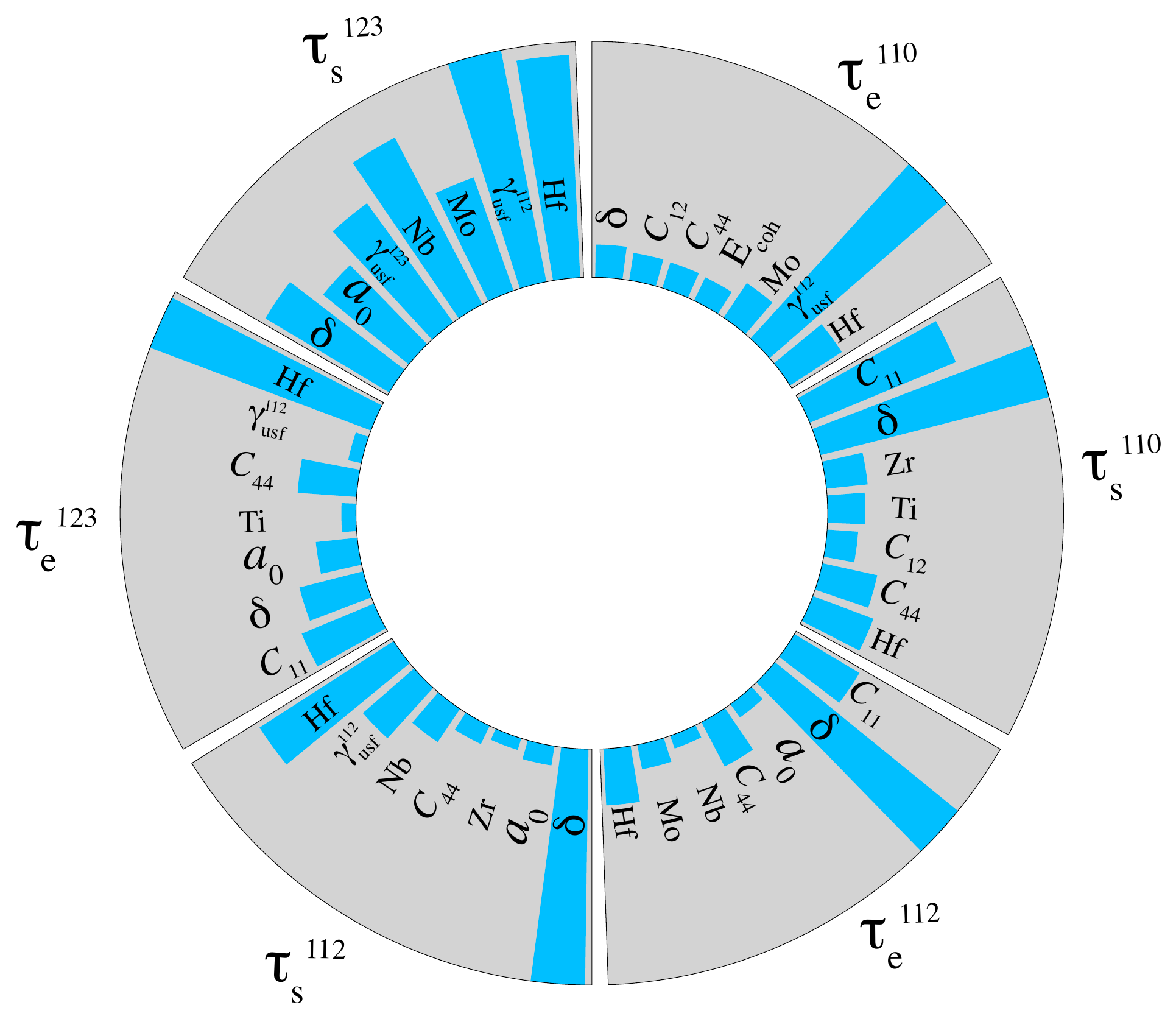}
\caption{Feature importance analysis for fundamental properties of 12 RMPEAs and weighted fractions of elements in the alloys, evaluated for six target variables (LSR values for edge and screw dislocations on three slip planes). Each sector in the pie chart corresponds to a target variable, with bars indicating feature importance ranked by the mean accuracy decrease score. The height of each bar is normalized to the bar with the most feature importance within each sector.}
\label{fig:11}
\end{figure}

The important features for six target variables (LSR values for edge and screw dislocations on three slip planes) are depicted in six sectors in the circular bar chart, respectively (\autoref{fig:11}). 
Feature importance is quantified via permutation analysis, with the mean accuracy decrease score measuring each feature's influence on the target variable. The most important features, i.e., those with the highest mean accuracy decrease score, are displayed in the chart, with the height of each bar representing the magnitude of the importance. 

\autoref{fig:11} reveals that the dominant feature for $\tau^{110}_\text{e}$ and $\tau^{123}_\text{s}$ is $\gamma^{112}_\text{usf}$; for $\tau^{110}_\text{s}$, $\tau^{112}_\text{e}$, and $\tau^{112}_\text{s}$, it is $\delta$; for $\tau^{123}_\text{e}$, it is the element Hf. When examining the influence of elemental compositions in alloys, the presence of element Hf is a key factor affecting all LSR values. Similarly, Mo or Nb also plays an important role, impacting all LSR values except $\tau^{110}_\text{s}$ and $\tau^{123}_\text{e}$. In terms of the fundamental properties, the elastic constant $C_{44}$ is critical for nearly all LSR values, with the exception of $\tau^{123}_\text{s}$. The lattice constant $a_0$ also directly influences most LSR values, except $\tau^{110}_\text{e}$ and $\tau^{110}_\text{s}$. Interestingly, $\delta$ proves to be a decisive factor across all LSR values.

The elastic constant $C_{44}$ is a direct measure of the material's resistance to shear deformation. At the atomic scale, the stress required to move a dislocation (the Peierls stress in pure metals or the LSR in RMPEAs) is intrinsically linked to the shear modulus. A higher $C_{44}$ indicates a stiffer lattice with stronger atomic bonds in the shear direction. This increased stiffness leads to a higher energy barrier for dislocation glide, as the dislocation core energy scales with the shear modulus \citep{hirth1982theory}, and an increased ISS, which is the theoretical upper limit of shear stress the lattice can sustain before instability. \autoref{tbl:mech_sum} shows a roughly positive correlation between $C_{44}$ and $\tau_{\text{iss}}$. Therefore, $C_{44}$ serves as a fundamental descriptor of the intrinsic ``hardness'' of the slip plane, making it an important governing parameter for LSR.

LD in RMPEAs arises from the atomic size mismatch among constituent elements. The LD coefficient $\delta$, quantified via the radial distribution function, measures the degree of this local atomic disorder. Its dominant influence on LSR can be understood through several interconnected mechanisms. LD creates a spatially fluctuating potential energy field for dislocations, resulting in local variations of the USFE and the Peierls barrier. Dislocations must navigate these ``rough'' energy landscapes, encountering local pinning sites, thereby influencing the average stress required for motion (i.e., the LSR). Moreover, the heterogeneous local atomic environment can non-uniformly distort the dislocation core. For screw dislocations in BCC RMPEAs, which have a non-planar, spread core, severe LD can further localize or modify the core structure, affecting the dislocation mobility. This is reflected in \autoref{fig:9}(a), where severe LD reduces the screw-to-edge LSR ratio, indicating a pronounced effect on screw dislocation mobility. Additionally, a large LD $\delta$ leads to a significant increase in the coefficient of variation (COV) of the LSR (\autoref{fig:6}), confirming that LD is the primary source of statistical scatter in LSR values. The fluctuating atomic environments sampled by dislocations in different locations lead to a wide range of local resistance values, with $\delta$ quantifying the amplitude of these fluctuations. Furthermore, LD differentially affects slip systems. \autoref{fig:9}(b-c) show that severe LD enhances the LSR ratios of $\{110\}$-to-$\{112\}$ and $\{110\}$-to-$\{123\}$, promoting activity on higher-index planes. This is because LD breaks the symmetry of the crystal field, altering the relative energy barriers on different planes and thereby changing the anisotropy of plastic slip.

Hf content significantly influences LD due to its large atomic radius. Hf-rich compositions tend to enhance local strain fields and modify the stacking fault energies, thereby affecting dislocation core structures and the LSR values across multiple slip systems. In addition, the peak value of the GSFE curve, the USFE, is related to the energy barrier for dislocation glide. A higher USFE generally corresponds to greater lattice resistance to slip, making it a dominant factor of LSR in systems where dislocation motion is governed by intrinsic bond-breaking processes.

In addition, the LSR may also depend on the changes in the dislocation elastic strain energy and the core energy as it moves one lattice periodicity. The dislocation core and elastic strain energies comprise its formation energy and these generally depend on screw/edge character, Burgers vector, elastic stiffness and degree of elastic anisotropy. In an RMPEA, the dislocation formation energy can vary significantly with location and among the slip planes. In the present atomistic approach, the fully relaxed dislocation configurations are used as initial states to compute LSR. The LSR calculation inherently captures the changes in the core structure within the local atomic environment and its associated strain energy as it moves through the lattice. Consequently, the calculated LSR values already reflect all contributions to the total energies, including changes in the core energy and line energy as it moves. While isolation of the formation energy of each dislocation studied for LSR and its connection to the LSR is interesting effort. we believe it is a great idea for a future study.

\subsection{Yield strength model versus LSR}
\label{sec:discussion}

\subsubsection{Existing experimental data}
%%%Table4
\begin{table*}[]
\rotatebox{90}{
\begin{minipage}{\textheight} % Keep the table aligned and inside page margins
\caption{The tensile properties of 7 RMPEAs reported in the literature. Grain size: $d$; Loading temperature: $T$; Strain rate: $\dot{\varepsilon}$; Yield strength: $\sigma_{\mathrm{Y}}$.}
\label{tbl:tension}
\centering
\resizebox{\textwidth}{!}{
\begin{tabular}{ccccc}
\hline
Composition & $d$ ($\mu\mathrm{m}$) & $T$ (K) & $\dot{\varepsilon}$ (s$^{-1}$) & $\sigma_{\mathrm{Y}}$ (MPa)  \\ \hline
NbTaTi \citep{wang2019intermetallics} & 341 & 300 & $5 \times 10^{-4}$ & 478 \\ \hline
HfNbTa \citep{wang2019intermetallics} & 215 & 300 & $5 \times 10^{-4}$ & 847 \\ 
HfNbTa \citep{sun2023msea} & 17 & 300 & $1 \times 10^{-3}$ & 989 \\ \hline
NbTiZr \citep{he2023scripta} & 0.208/0.231/0.91/1.6/4.0/7.1/11.4/21.6 & 300 & $8.3 \times 10^{-4}$ & 1045/1042/826/876/768/754/768/745 \\ \hline
HfNbTi \citep{he2023scripta} & 0.512/1.253/2.215/2.428/7.13/8.86/19.24 & 300 & $8.3 \times 10^{-4}$ & 909/783/763/794/739/714/730 \\ \hline
HfTaTi \citep{he2023scripta} & 0.788/1.121/1.765/3.474/5.658/12.125 & 300 & $8.3 \times 10^{-4}$ & 976/928/926/862/828/808 \\ \hline
HfNbTaTi \citep{wang2019intermetallics} & 340 & 300 & $5 \times 10^{-4}$ & 663 \\
HfNbTaTi \citep{sun2023msea} & 27 & 300 & $1 \times 10^{-3}$ & 843 \\
HfNbTaTi \citep{he2023scripta} & 0.344/0.511/1.603/2.076/4.399/6.908/9.43 & 300 & $8.3 \times 10^{-4}$ & 1114/1108/944/890/880/923/839 \\ \hline
HfNbTaTiZr \citep{mills2023acta} & 100 & 300/1073/1473 & $1 \times 10^{-2}$ & 964/444/30 \\
HfNbTaTiZr \citep{juan2016ml} & 38/81/128 & 300 & $1 \times 10^{-3}$ & 958/944/940 \\
HfNbTaTiZr \citep{chen2021scripta} & 40 & 673/773/823 & $1 \times 10^{-3}$ & 817/819/734 \\
HfNbTaTiZr \citep{chen2021scripta} & 40 & 573/673/723/773 & $2 \times 10^{-4}$ & 806/735/689/698 \\
HfNbTaTiZr \citep{chen2021scripta} & 40 & 573/673/723/773 & $5 \times 10^{-5}$ & 854/759/718/722 \\\hline
\end{tabular}
}
\end{minipage}
}
\end{table*}

The concept of an LSR is central to understanding the strength of RMPEAs; however, the LSR prevails at the nanoscale while strength is measured at the macroscale. To help bridge the length scale gap and connect the LSR to material strength, we collect as many experimental tensile yield-strength data as possible for the RMPEAs examined in our atomistic calculations. Here, we only focus on the tensile yield strength instead of compressive yield strength, since the tensile performance of an RMPEA more directly reflects real-world demands for fracture resistance, ductility, and high-temperature stability. As summarized in \autoref{tbl:tension}, these data span various RMPEAs (NbTaTi, HfNbTa, NbTiZr, HfNbTi, HfTaTi, HfNbTaTi, and HfNbTaTiZr), covering a wide range of grain sizes (from hundreds of nanometers to hundreds of micrometers), temperatures (from 300 K to 1473 K), and strain rates ($5 \times 10^{-5}$ s$^{-1}$ to $1 \times 10^{-2}$ s$^{-1}$). The measured tensile yield strengths range from tens of MPa to thousands of MPa. Moreover, for a given MPEA composition tested under the same conditions, increasing the grain size or the temperature generally leads to a decrease in the tensile yield strength.

\subsubsection{Comparison with analytical models}
In RMPEAs, dislocation motion is governed by the interplay of the thermal energy, applied stress, and LSR (analogous to the Peierls stress in pure metals). Thermally activated dislocation motion proceeds by overcoming an energy barrier, and the applied strain rate $\dot{\varepsilon}$ can be related to the absolute temperature $T$ and the activation enthalpy $\Delta H$ for the local dislocation motion via an Arrhenius-like equation \citep{lim2015ijp}:
\begin{equation}
\dot{\varepsilon} = \dot{\varepsilon}_0\, \mathrm{exp} \left(-\frac{\Delta H}{k_\mathrm{B}T}\right),
\label{equ:3}
\end{equation}
where $\dot{\varepsilon}_0$ and $k_\mathrm{B}$ are the reference strain rate and the Boltzmann constant, respectively. For BCC RMPEAs, $\dot{\varepsilon}_0$ is commonly taken as $10^4$ s$^{-1}$ \citep{ghafarollahi2022acta,he2023scripta}.
In this work, both edge and screw dislocations on the $\{110\}$, $\{112\}$ and $\{123\}$ slip planes, six types in total, are considered. These dislocations are labeled ($i = 1, 2, ..., 5, 6$), where (1) edge dislocations on the $\{110\}$ slip plane, (2) screw dislocations on the $\{110\}$ slip plane, (3) edge dislocations on the $\{112\}$ slip plane, (4) screw dislocations on the $\{112\}$ slip plane, (5) edge dislocations on the $\{123\}$ slip plane, and 6) screw dislocations on the $\{123\}$ slip plane. For either the elastic interaction or line tension activation enthalpy models, the activation enthalpy is generally written as \citep{kocks1975pms}:
\begin{equation}
\Delta H_i(\tau) = \Delta H_{0,i} \left(1- \left(\frac{\tau_i}{\tau_{\mathrm{lsr},i}} \right)^p \right)^q,
\label{equ:4}
\end{equation}
where $\Delta H_{0,i}$, $\tau_i$, and $\tau_{\mathrm{lsr},i}$ denote the average energy barrier for the local motion of the $i_\mathrm{th}$ dislocation, the applied resolved shear stress and the average LSR value of the $i_\mathrm{th}$ dislocation, respectively. The exponents $p$ and $q$ control the shape of the energy barrier and typically satisfy $0 \leqslant p \leqslant 1$ and $1 \leqslant q \leqslant 2$. For BCC RMPEAs, it is common to use $p = 1$ and $q = 3/2$ \citep{ghafarollahi2022acta,kubilay2021npjcm}.
By combining \autoref{equ:3} and \autoref{equ:4}, one can derive
\begin{equation}
\tau_i = \tau_{\mathrm{lsr},i} \left(1-\left(\frac{k_\mathrm{B}T}{\Delta H_{0,i}} \mathrm{ln}\frac{\dot{\varepsilon}_0}{\dot{\varepsilon}} \right)^{\frac{2}{3}} \right).
\label{equ:5}
\end{equation}

Experimental evidence indicates that, in BCC RMPEAs, both edge and screw dislocations on the $\{110\}$, $\{112\}$, and $\{123\}$ slip planes collectively contribute to the overall yield behavior \citep{wang2020science}. The yield stress can therefore be estimated using a Hall-Petch-type equation:
\begin{equation}
\sigma_\mathrm{Y} = K_\mathrm{HP}\, d^{-\frac{1}{2}} + M \sum_{i=1}^{6} \tau_i a_i,
\label{equ:6}
\end{equation}
where $K_\mathrm{HP}$ is the Hall--Petch coefficient, $d$ is the average grain size, $M$ is the Taylor factor (typically taken as 2.733 for BCC RMPEAs \citep{ghafarollahi2022acta,he2023scripta}), and $a_i$ represents the slip-system-dependent interaction coefficient ($i = 1,\dots,6$). 

\begin{figure}[htb!]
\centering
\includegraphics[width=0.9\linewidth]{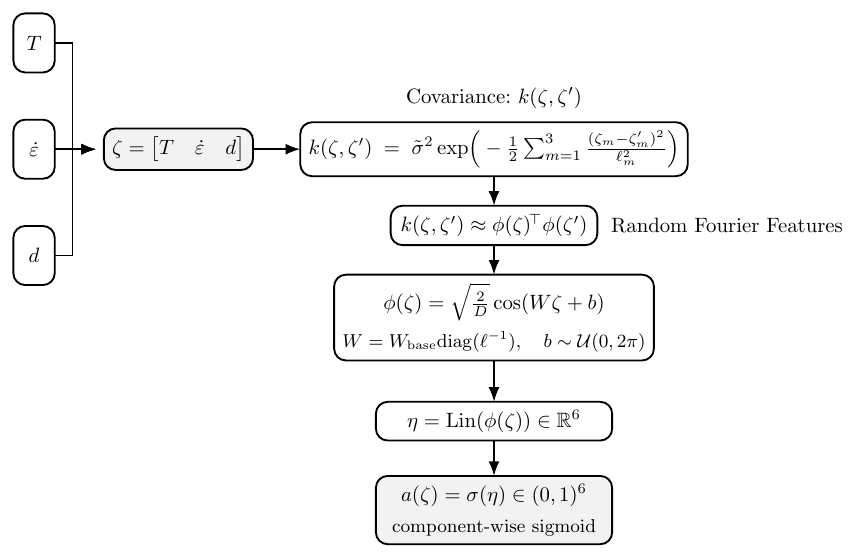}
\caption{Schematic illustrating the modeling of interaction coefficients $a_i$ as context-dependent quantities via Gaussian process regression. The notations $a(\zeta)$ and $a_i(\zeta)$ are used interchangeably, with $a(\zeta) = [a_i(\zeta)]$ defining the relationship. The overall input and output are marked in gray.}
\label{fig:ai_schematic}
\end{figure}

Unlike the conventional assumption of constant $a_i$, the interaction coefficients are modeled as context-dependent quantities derived from Gaussian process regression. This approach is superior because it allows the model to capture the complex, nonlinear dependence of dislocation interactions on evolving thermomechanical conditions, rather than being limited to a single averaged behavior. Specifically, the interaction coefficient $a_i(\zeta)$ is constructed as a 6-element vector $a(\zeta)$, which depends on the local context vector $\zeta=[T, \dot{\varepsilon}, d]$, capturing the effects of temperature, strain rate, and grain size on dislocation interactions. A schematic illustration of this construction is provided in \autoref{fig:ai_schematic}.

Here, the input context vector is $\zeta=[T,\dot{\varepsilon},d]$. 
The Gaussian covariance kernel is defined as:
\begin{equation}
k(\zeta, \zeta') = \tilde{\sigma}^2 \exp\left(-\frac{1}{2} \sum_{m=1}^{3} \frac{(\zeta_m - \zeta_m')^2}{\ell_m^2} \right),
\label{equ:GPR_kernel}
\end{equation}
where $\tilde{\sigma}^2$ is the kernel variance, and $\ell=(\ell_1,\ell_2,\ell_3)=(\ell_T,\ell_{\dot\varepsilon},\ell_d)$ are the Automatic Relevance Determination (ARD) length scales, corresponding to temperature $T$, strain rate $\dot{\varepsilon}$, and grain size $d$, respectively. $\zeta$ and $\zeta'$ represent two distinct context vectors, and $k(\zeta,\zeta')$ quantifies the similarity between the corresponding material and loading conditions ($T$, $\dot{\varepsilon}$, $d$). The kernel $k(\zeta,\zeta')$ can be approximated using random Fourier features (RFF):
\begin{equation}
k(\zeta,\zeta')\approx \phi(\zeta)^{\!\top}\phi(\zeta'),
\label{equ:RFF_approx}
\end{equation}
where $\phi(\zeta)=\sqrt{2/D}\cos(W\zeta+b)$. The matrix $W=W_{\text{base}}\operatorname{diag}(\ell^{-1})$ is constructed such that $W_{\text{base}}\in\mathbb{R}^{D\times 3}$ has independent and identically distributed entries sampled from $\mathcal{N}(0,1)$, and at the $d_\text{th}$ row $W_d\sim\mathcal{N}(0,\operatorname{diag}(\ell^{-2}))$ represents sampled frequencies, drawn from a multivariate normal distribution with zero mean and covariance $\operatorname{diag}(\ell^{-2})$. The scaling factor $\operatorname{diag}(\ell^{-1})$ applies ARD weighting to the sampled frequencies.
The vector $b\sim\mathcal{U}(0,2\pi)$ contains random phases drawn from the uniform distribution over $[0,2\pi)$.
The ARD length scales $\ell$ control the sensitivity of the model to each input dimension: a small $\ell_m$ increases sensitivity to the $m_\text{th}$ input, while a larger $\ell_m$ reduces it. A linear transformation $\eta = \rm Lin(\phi(\zeta))$ is then applied to $\phi(\zeta)$, producing six independent components corresponding to the slip systems. Finally, the full interaction vector $a(\zeta)=[a_i(\zeta)]$ is obtained by applying a sigmoid activation to bound the components within $(0,1)$.

With this extension, the theoretical expression for the yield strength of a BCC RMPEA becomes
\begin{equation}
\sigma_\mathrm{Y} = K_\mathrm{HP}\, d^{-\tfrac{1}{2}}
+ M \sum_{i=1}^{6} \tau_{\mathrm{lsr},i}
\left[1 - \left(\frac{k_\mathrm{B}T}{\Delta H_{0,i}}
\ln \frac{\dot{\varepsilon}_0}{\dot{\varepsilon}}\right)^{\tfrac{2}{3}} \right]\,a_i(\zeta).
\label{equ:YS}
\end{equation}
Here, LSR ($\tau_{\mathrm{lsr},i}$) is the athermal stress required to move a dislocation at 0 K. It sets the upper limit of strength. A higher LSR indicates a higher potential strengthening capacity. The term in square brackets $[1 - (\ldots)^{2/3}]$ represents the thermal assistance. As temperature $T$ increases, this term decreases, reducing the effective stress $\tau_i$ needed from the applied load to move the dislocation, even though $\tau_{\text{lsr},i}$ itself remains constant. The experimental data in \autoref{tbl:tension} shows that strength decreases with temperature, which is the classic effect of thermal activation overwhelming the athermal lattice friction. While a high LSR is desirable as it raises the overall strength ceiling, the thermal activation factor diminishes more rapidly with temperature. Therefore, the observed decrease in $\sigma_{\rm Y}$ with $T$ in \autoref{tbl:tension} is consistent with a high LSR.

By fitting \autoref{equ:YS} to combined experimental data ($\sigma_\mathrm{Y}$, $d$, $T$, $\dot{\varepsilon}$) and atomistic inputs ($\tau_{\mathrm{lsr},i}$), the model parameters ($K_\mathrm{HP}$, $\Delta H_{0,i}$, and the kernelized distribution $a_i(\zeta)$) can be identified. Once calibrated, the model provides predictive capability across a wide range of grain sizes, temperatures, and strain rates, enabling direct comparison with experimental yield stresses. The temperature dependence within this model is explicitly incorporated through two primary mechanisms. First, the thermally activated term in \autoref{equ:5} captures the fundamental reduction of the applied resolved shear stress $\tau_i$ with increasing temperature $T$. Second, the model accounts for the influence of thermal conditions on dislocation interactions through the context-dependent coefficients $a_i(\zeta)$, which are formulated as functions of temperature $T$, alongside strain rate $\dot{\varepsilon}$ and grain size $d$.
%The fitting procedures consider parameter ranges of 150–300 MPa for the Hall-Petch coefficient $K_\mathrm{HP}$ and 1.5–6.0 eV for the activation enthalpy $\Delta H_{0,i}$.

Our thermally activated framework shares a fundamental physical basis with models developed for screw-dislocation-controlled plasticity, such as those advanced by Curtin et al. \citep{maresca2020acta2,ghafarollahi2022acta}. Both frameworks describe yield as a stress-aided, thermally activated process following an Arrhenius law, where the energy barrier for dislocation motion is reduced by the applied stress. The key distinction lies in the specific mechanisms and dislocations considered. The framework developed by Curtin et al. \citep{maresca2020acta2,ghafarollahi2022acta} is built upon the concept of a spontaneously kinked screw dislocation in random alloys. The alloy strength is controlled by the glide of these intrinsic kinks and the unpinning of cross-kinks on  $\{110\}$ planes. In contrast, our model explicitly incorporates the contributions of both edge and screw dislocations on the three predominant slip planes ($\{110\}$, $\{112\}$, $\{123\}$). The activation enthalpy in our formulation is directly linked to the LSR, a statistically distributed quantity that captures the atomistic-scale resistance arising from chemical disorder and lattice distortion. This formulation allows our model to account for scenarios where edge dislocations on higher-index planes contribute significantly to yield, which is particularly relevant in chemically complex RMPEAs.

%%%Fig13
\begin{figure}[htb!]
\centering
\includegraphics[scale=0.5,clip]{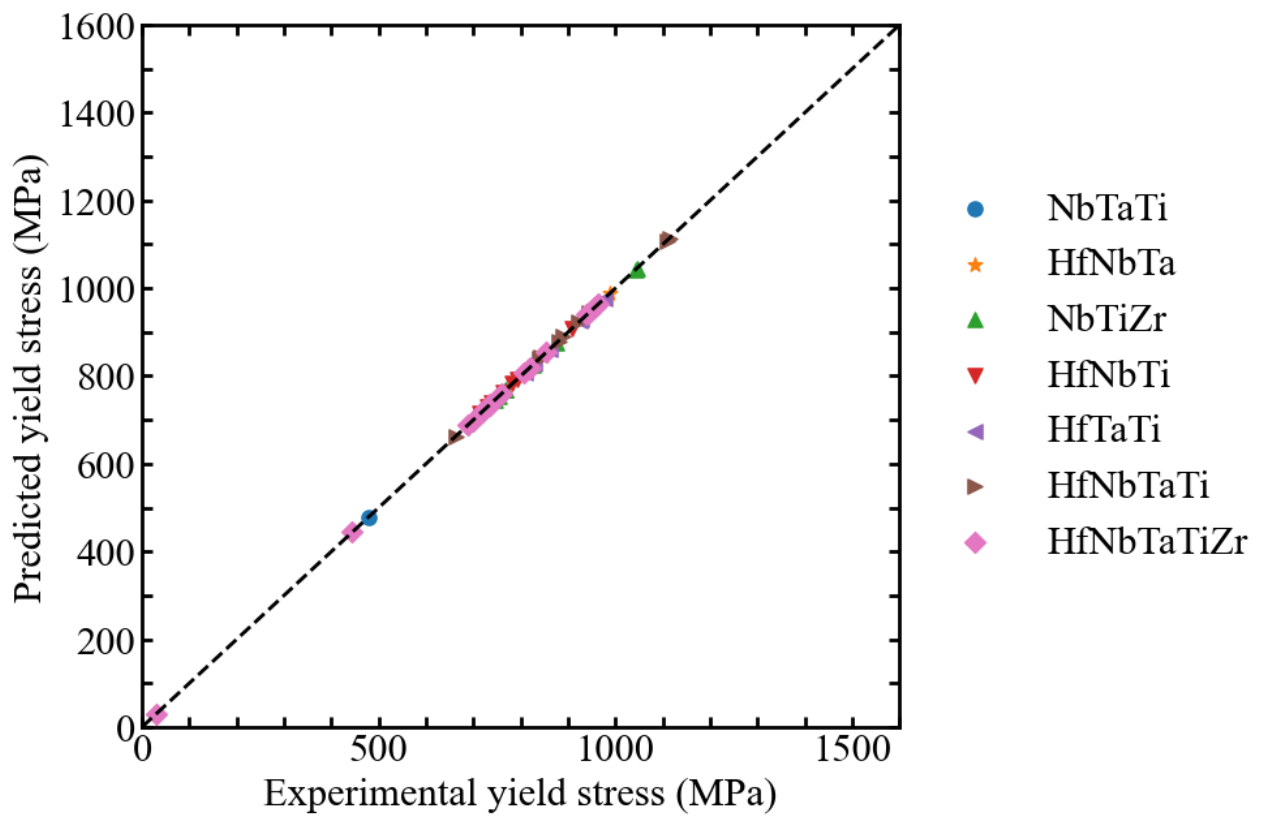}
\caption{Comparison of theoretical yield-strength predictions and experimental measurements for seven RMPEAs.}
\label{fig:comparison}
\end{figure}

\autoref{fig:comparison} compares the theoretical predictions from \autoref{equ:YS} with experimentally measured yield strengths for seven RMPEAs. Although the experimental data were obtained from different studies in different laboratories, the predicted values agree well with the measurements. Here, all experimental yield strength data listed in \autoref{tbl:tension} are used to calibrate the model parameters. This includes fitting the alloy-specific Hall--Petch coefficient $K_{\mathrm{HP}}$, the six average activation enthalpies $\Delta H_{0,i}$ for the considered dislocation types, and the hyperparameters of the context-dependent interaction kernel. \autoref{equ:YS} is fitted to the entire dataset by minimizing the mean squared error. Specifically, the loss function is defined as the sum of squared differences between the experimentally measured yield strengths and the model predictions. Given the relatively limited number of available experimental data points for the specific RMPEA compositions studied here, we use the full dataset for calibration to ensure robust parameter estimation, acknowledging that this precludes a formal hold-out validation. The model is strongly physics-informed, and the fitted parameters $\Delta H_{0,i}$ and $K_{\mathrm{HP}}$, listed in Tables~\ref{tbl:Delta_H} and~\ref{tbl:K_HP} respectively, fall within physically plausible ranges \citep{he2023scripta}. We view this framework as a calibrated predictive tool for interpolation within the studied conditions. Future work with more extensive datasets will enable rigorous out-of-sample testing.

As shown in \autoref{fig:comparison_single}, when the six interaction coefficients $a_i$ in \autoref{equ:YS} are set to a single parameter instead of a context-dependent function, there is a significant decrease in prediction accuracy, although the results remain in reasonable agreement with experimental data. This outcome is understandable, since the interactions between different types of dislocations on different glide planes in different alloy systems under different loading conditions are too complex to be adequately captured by six constant $a_i$ parameters. Alternatively, the context-dependent formulation via Gaussian processes ensures that $a_i(\zeta)$ varies smoothly with $T$, $\dot{\varepsilon}$, and $d$, reducing overfitting risk.

Based on the data in \autoref{tbl:Delta_H} and \autoref{tbl:lsr_sum}, a complex and non-uniform correlation is observed between the activation enthalpy ($\Delta H_0$) and the LSR values across different dislocation types and slip planes. This trend stems from their common origin in the intrinsic lattice friction and the core structure of dislocations. Higher $\Delta H_0$ values indicate greater energy barriers for dislocation motion, which typically translate to higher LSR required to initiate plastic deformation. For instance, in NbTaTi alloy, the screw dislocation on the $\{110\}$ plane exhibits both a high $\Delta H_0$ (4.3317 eV) and a correspondingly high LSR (1731.11 MPa), confirming its significant resistance to slip. Conversely, lower $\Delta H_0$ values, such as for the edge dislocation on the $\{123\}$ plane in NbTaTi alloy (1.9835 eV), generally align with a lower LSR value (340.94 MPa), indicating enhanced dislocation mobility. However, this correlation is not strictly linear across all systems, as evidenced by HfNbTaTiZr alloy where edge dislocations on $\{112\}$ planes show the highest $\Delta H_0$ (5.9575 eV) but only moderate LSR (659.58 MPa), reflecting the complex influence of local chemical environments, solid solution strengthening effects, and varying dislocation core responses to different alloy compositions in these RMPEAs.

The theoretically fitted Hall-Petch coefficients $K_\mathrm{HP}$ for these alloys, listed in \autoref{tbl:K_HP}, range from 150.30 to 186.23 $\mathrm{MPa} \cdot \mu\mathrm{m}^{-0.5}$, which is consistent with those obtained from experiments \citep{he2023scripta}. Notably, there is no direct correlation between $K_\mathrm{HP}$ and the average intrinsic slip resistance parameters ($\Delta H_0$ or LSR) for any specific dislocation type. For example, HfTaTi has the highest $K_\mathrm{HP}$ of 186.23 $\mathrm{MPa} \cdot \mu\mathrm{m}^{-0.5}$ but exhibits moderate $\Delta H_0$ (ranging from 1.7027 to 3.2618 eV for edge dislocations and from 2.1156 to 4.0068 eV for screw dislocations) and moderate average LSR values (ranging from 549.66 to 912.06 MPa for edge dislocations and from 740.18 to 1013.25 MPa for screw dislocations). This suggests that when intrinsic lattice friction is not overwhelmingly high, grain boundaries may become relatively more effective barriers to dislocation motion, thereby enhancing the Hall-Petch strengthening contribution. The variability in $K_\mathrm{HP}$ is thus likely governed by complex grain boundary-related factors, such as dislocation nucleation, transmission, and the anisotropy of slip systems, rather than being solely determined by the intrinsic properties of dislocation glide.

This finding demonstrates that the thermally activated, dislocation-based model captures the yield behavior of BCC RMPEAs over a wide range of grain sizes, temperatures, and strain rates. In contrast to BCC pure metals, where screw dislocations predominantly control plastic deformation \citep{lim2015ijp}, both edge and screw dislocations on the most common slip planes ($\{110\}$, $\{112\}$, $\{123\}$) govern the deformation mechanisms in the BCC RMPEAs. Consequently, the proposed model provides a reliable framework for understanding and predicting the yield strengths of BCC RMPEAs, facilitating their design and optimization in various applications. While it is recognized that conventional EAM potentials have limitations in describing screw dislocation cores in some BCC metals, the interatomic potential employed in this work \citep{mubassira2025cms} is specifically developed for RMPEAs and validated against density functional theory (DFT) data for stacking fault energies and elastic constants, thereby providing a reliable atomistic foundation for the calculated LSRs. More importantly, the thermally‑activated framework expressed in \autoref{equ:YS} is not tied to a particular interatomic potential; it is formulated in terms of generic dislocation‑scale parameters (LSRs, activation enthalpies, and interaction coefficients). If more accurate LSR values are obtained, whether from DFT‑based nudged‑elastic‑band calculations or from improved machine‑learning potentials, they can be directly used in \autoref{equ:YS} without changing the model structure. This makes the approach transferable to other alloy systems and to higher‑fidelity data, provided the underlying dislocation‑mediated yield mechanism remains valid.

\subsection{Implications for alloy design}
The insights gained from this study establish a clear, mechanism-informed pathway for designing next-generation RMPEAs with superior mechanical performance. The central design philosophy is to maximize the alloy's intrinsic resistance to dislocation glide while maintaining sufficient pathways for plasticity. This is achieved by engineering a high and spatially uniform LSR through strategic chemical design, which directly targets the key descriptors identified in our ML analysis: the shear modulus ($C_{44}$), the lattice distortion coefficient ($\delta$), and the weighted fractions of specific elements.

Concretely, this leads to a multi-pronged strategy. First, severe LD should be maximized by incorporating elements with large atomic size mismatch (e.g., Hf and Nb) to raise the plasticity. Second, this can be strategically combined with a high fraction of HCP elements for ductility and strong BCC elements like Mo and Ta to elevate key intrinsic properties (e.g., $C_{44}$, USFE, ISS), creating a BCC matrix with high LSRs and then a high intrinsic strength ceiling. Third, although this study focuses on single-phase BCC alloys, thermally stable coherent or semi-coherent secondary phases (e.g., B2, Laves) can be introduced to provide additional athermal strengthening via precipitation mechanisms and to pin grain boundaries, thereby supplementing the matrix strength with Hall-Petch grain refinement.

To efficiently implement this strategy, the atomistically-informed, data-driven framework developed in this work provides a practical tool for composition selection and property prediction. Our model creates a powerful, closed-loop design paradigm. Designers can first use the identified descriptors ($C_{44}$, $\delta$, elemental fractions) as quantitative targets for high-throughput computational screening of candidate compositions. For the most promising candidates, the essential atomistic input parameters (e.g., the LSR values) can be determined via the established simulation protocol. These parameters are then fed into the thermally activated yield strength model to predict macroscopic performance across a wide range of thermomechanical conditions (\(T\), \(\dot{\varepsilon}\), \(d\)). Finally, experimental validation of synthesized alloys provides new data to iteratively refine the model's accuracy, creating a virtuous cycle that accelerates the discovery of alloys where exceptional high-temperature strength is achieved without sacrificing ductility.

\section{Conclusions}
In this work, we perform atomistic simulations to calculate the local slip resistances (LSRs) of edge and screw dislocations in 12 RMPEAs on the three most common slip planes ($\{110\}$, $\{112\}$, $\{123\}$), as well as to determine the fundamental properties of these multicomponent alloys. The machine learning (ML) methods are utilized to analyze how these LSR values relate to the fundamental properties of RMPEAs. Subsequently, a thermally activated, dislocation-based model is developed to predict the yield stress of RMPEAs across a broad range of grain sizes, temperatures, and strain rates. The main findings are summarized below:

\begin{itemize}  

\item A high fraction of HCP elements (exceeding 50\%) in the composition lowers both the unstable stacking fault energy (USFE) and the ideal shear strength (ISS) in BCC RMPEAs. Moreover, increased elastic anisotropy, as indicated by a rising Zener ratio, also reduces USFE and ISS. 

\item The LSRs in BCC RMPEAs are governed by multiple factors. Increases in the Zener ratio, along with decreases in USFE and ISS, lead to reduced LSR values. Additionally, Hf-, Ti- or Zr-containing alloys with more than 50\% HCP elements exhibit low LSR values for screw dislocations on all three slip planes.

\item Severe lattice distortion (LD) with $\delta > 0.03$ diminishes the screw-to-edge LSR ratios but enhances the \(\{110\}\)-to-\(\{112\}\) and \(\{110\}\)-to-\(\{123\}\) LSR ratios, thereby affecting plasticity anisotropy in these alloys.

\item The presence of Hf in the alloy composition strongly affects the LSR values for both edge and screw dislocations across all three slip planes, while Mo or Nb impacts nearly all LSR values except for $\tau^{110}_{\rm s}$ and $\tau^{123}_{\rm e}$.

\item The elastic constant $C_{44}$, LD, and lattice constant $a_0$, consistently emerge as important parameters influencing the LSR values of dislocations.

\item The macroscopic yield strength of a BCC RMPEA can be effectively predicted by a thermally activated, dislocation-based model that accounts for both edge and screw dislocations on the most common slip planes ($\{110\}$, $\{112\}$, $\{123\}$). This confirms that various dislocation types on both low- and high-index slip planes collectively determine the yield behaviors of BCC RMPEAs.

\end{itemize}

\section*{Data and code availability}
The data and code supporting this work are publicly available via \url{https://github.com/wrj2018/LSR_2024}.

\section*{Acknowledgements}
Use was made of computational facilities purchased with funds from the National Science Foundation (No. CNS-1725797) and administered by the Center for Scientific Computing (CSC). The CSC was supported by the California NanoSystems Institute and the Materials Research Science and Engineering Center (MRSEC; NSF DMR 2308708) at UC Santa Barbara. S.X.\ acknowledges the support of the U.S.\ National Science Foundation (DMREF-2522655). X.Y.\ acknowledges support from the National Natural Science Foundation of China (No.\ 12232006). W.J.\ acknowledges support from the National Natural Science Foundation of China (No. U2441214).

\hspace{10pt}

\setcounter{figure}{0}
\setcounter{table}{0}

\clearpage
\appendix

\section{}
%%%TableA1
\begin{table*}[htp]
\caption{The fitting average energy barrier $\Delta H_{0,i}$ (Unit: eV) for the local motion of the $i_\mathrm{th}$ dislocation for \autoref{equ:YS}. These dislocations are labeled ($i = 1, 2, ..., 5, 6$), where (1) edge dislocations on the $\{110\}$ slip plane, (2) screw dislocations on the $\{110\}$ slip plane, (3) edge dislocations on the $\{112\}$ slip plane, (4) screw dislocations on the $\{112\}$ slip plane, (5) edge dislocations on the $\{123\}$ slip plane, and 6) screw dislocations on the $\{123\}$ slip plane.}\label{tbl:Delta_H}
\centering
\begin{tabular}{ccccccc}
\hline
Composition & $\Delta H_{0,1}$ & $\Delta H_{0,2}$ & $\Delta H_{0,3}$ & $\Delta H_{0,4}$ & $\Delta H_{0,5}$ & $\Delta H_{0,6}$ \\ \hline
NbTaTi & 4.0704 & 4.3317 & 2.7891 & 3.6344 & 1.9835 & 2.7206 \\ \hline
HfNbTa & 3.2532 & 2.9563 & 3.8605 & 2.8993 & 3.5755 & 2.3528 \\ \hline
NbTiZr & 4.2213 & 5.6719 & 3.2275 & 3.6783 & 2.9467 & 3.3532 \\ \hline
HfNbTi & 5.2102 & 1.7932 & 2.2484 & 2.3665 & 4.8176 & 3.2876 \\ \hline
HfTaTi & 2.6356 & 2.1156 & 1.7027 & 2.3867 & 3.2618 & 4.0068 \\ \hline
HfNbTaTi & 2.6496 & 3.5209 & 3.2088 & 3.1779 & 2.2332 & 2.7168 \\ \hline
HfNbTaTiZr & 5.3470 & 3.0537 & 5.9575 & 1.6802 & 2.9012 & 1.6174 \\ \hline
\end{tabular}
\end{table*}

%%%TableA2
\begin{table*}[htp]
\caption{The fitting Hall-Petch coefficient $K_\mathrm{HP}$ (Unit: $\mathrm{MPa} \cdot \mu\mathrm{m}^{-0.5}$) for \autoref{equ:YS}.}\label{tbl:K_HP}
\centering
\begin{tabular}{cccccccc}
\hline
 & NbTaTi & HfNbTa & NbTiZr & HfNbTi & HfTaTi & HfNbTaTi & HfNbTaTiZr  \\ \hline
$K_\mathrm{HP}$ & 171.16 & 177.00 & 150.30 & 165.24 & 186.23 & 154.29 & 160.98 \\ \hline
\end{tabular}
\end{table*}

%%%FigureA1
\begin{figure*}[htb!]
\centering
\includegraphics[width=0.75\linewidth]{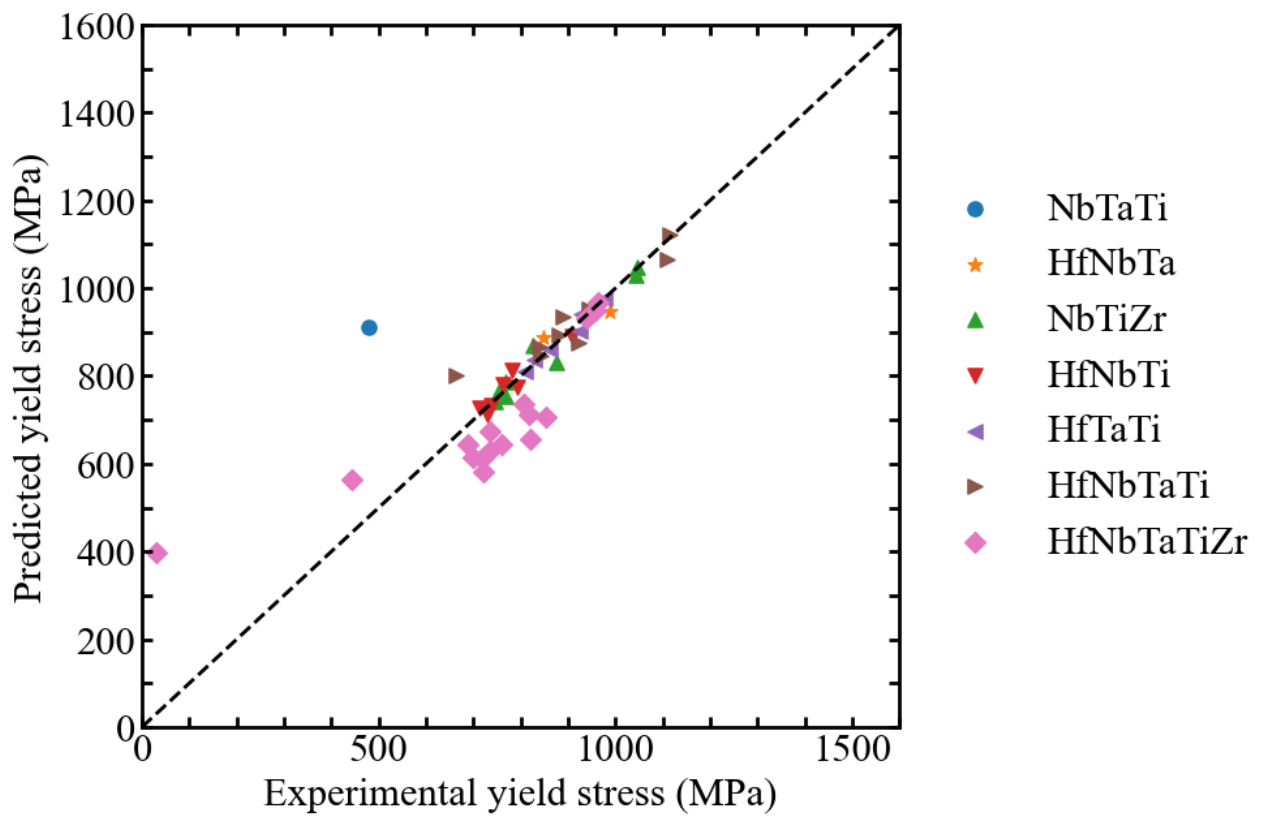}
\caption{Comparison of theoretical yield-strength predictions and experimental measurements for seven
RMPEAs when the six interaction coefficients $a_i$ in \autoref{equ:YS} are set to a single parameter instead of a context-dependent function.}
\label{fig:comparison_single}
\end{figure*}

\bibliographystyle{plainnat}
\bibliography{ref}

\end{document}